\documentclass[lettersize,journal]{IEEEtran}

\usepackage{amsmath,amssymb,amsfonts}
\usepackage{xfrac}
\usepackage{array}
\usepackage{bm}
\usepackage{algorithm}
\usepackage{algpseudocode}

\usepackage[table,dvipsnames]{xcolor}
\usepackage{graphicx}
\usepackage{booktabs}
\usepackage{multirow}
\usepackage{tabularx}

\usepackage{hyperref}
\usepackage[capitalise]{cleveref}
\usepackage{xr}
\newcommand{\crefnobox}[1]{\cref{#1}}  

\usepackage{csquotes}
\usepackage{orcidlink}
\usepackage{pifont}

\usepackage{footnotehyper}
\makesavenoteenv{tabular}
\makesavenoteenv{table}
\usepackage[nolist]{acronym}
\usepackage{caption}
\usepackage{subcaption}
\usepackage{svg}
\usepackage{varwidth}
\usepackage{tikz}
\usepackage{pgf}
\usepackage{pgfplots}
\usepackage{pgfplotstable}
\usetikzlibrary{positioning,backgrounds,fit,calc,patterns,arrows.meta}
\usetikzlibrary{shapes.geometric}
\usetikzlibrary{shapes.arrows}
\usetikzlibrary{arrows.meta}

\usepackage[subtle]{savetrees}
\usepackage{microtype}

\definecolor{uhhblue}{RGB}{0,156,209}
\definecolor{uhhgreen}{RGB}{66, 178, 60}
\definecolor{uhhred}{RGB}{226,0,26}
\definecolor{uhhblack}{RGB}{0,0,0}
\definecolor{uhhstone}{RGB}{59,81,91}
\newcommand{\cmark}{\color{uhhgreen}{\ding{52}}}
\newcommand{\xmark}{\color{uhhred}{\ding{54}}}

\newcommand{\B}[1]{\textbf{#1}}
\newcommand{\G}[1]{\textcolor{ForestGreen}{#1}}
\newcommand{\R}[1]{\textcolor{red}{#1}}
\newcommand{\U}[1]{\underline{#1}}

\newcommand{\E}{\mathbb{E}}  %

\newcommand{\auC}{s}  %
\newcommand{\auE}{\hat{s}}  %
\newcommand{\auD}{y}  %
\newcommand{\auIE}{z}

\newcommand{\spec}{X}
\newcommand{\specC}{S}
\newcommand{\specD}{Y}
\newcommand{\specE}{\hat{\specC}}
\newcommand{\specIE}{Z}

\renewcommand{\d}{\mathrm{d}}  %
\newcommand{\loss}{\mathcal{L}}

\newcommand{\flow}{\phi}
\newcommand{\vel}{u}
\newcommand{\lvel}{v_\theta}

\newcommand{\pdata}{p_{\rm{data}}}

\DeclareMathOperator{\STFT}{STFT}
\DeclareMathOperator{\iSTFT}{iSTFT}
\renewcommand{\time}{t}
\newcommand{\dtime}{\tau}
\newcommand{\Gaussian}{\mathcal{N}}

\newcommand{\lat}{\ell}
\newcommand{\latalg}{\lat_\text{alg}}
\newcommand{\lattot}{\lat_\text{tot}}
\newcommand{\latproc}{\lat_\text{proc}}

\newcommand{\lay}{l} %

\newcommand{\winlenana}{W_\text{ana}}
\newcommand{\winlensynth}{W_\text{syn}}
\newcommand{\hopana}{H_\text{ana}}
\newcommand{\hopsynth}{H_\text{syn}}

\newcommand{\nan}{\texttt{NaN}}

\newcommand{\upsample}{\uparrow\!\uparrow}
\newcommand{\downsample}{\downarrow\!\downarrow}

\newcolumntype{L}[1]{%
  >{\begin{lrbox}{\nowrapbox}\begin{minipage}{#1}\raggedright}%
  l%
  <{\end{minipage}\end{lrbox}\usebox{\nowrapbox}}%
}

\begin{acronym}
\acro{sgm}[SGM]{score-based generative model}
\acro{snr}[SNR]{signal-to-noise ratio}
\acro{stft}[STFT]{short-time Fourier transform}
\acro{pr}[PR]{phase retrieval}
\acro{istft}[iSTFT]{inverse short-time Fourier transform}
\acro{sde}[SDE]{stochastic differential equation}
\acro{ode}[ODE]{ordinary differential equation}
\acro{pesq}[PESQ]{Perceptual Evaluation of Speech Quality}
\acro{se}[SE]{speech enhancement}
\acro{tf}[T-F]{time-frequency}
\acro{rir}[RIR]{room impulse response}
\acro{snr}[SNR]{signal-to-noise ratio}
\acro{bwe}[BWE]{bandwidth extension}
\acro{lstm}[LSTM]{long short-term memory}
\acro{polqa}[POLQA]{Perceptual Objective Listening Quality Analysis}
\acro{sdr}[SDR]{signal-to-distortion ratio}
\acro{lsd}[LSD]{log-spectral distance}
\acro{sisdr}[SI-SDR]{scale invariant signal-to-distortion ratio}
\acro{estoi}[ESTOI]{Extended Short-Term Objective Intelligibility}
\acro{drr}[DRR]{direct-to-reverberant ratio}
\acro{nfe}[NFE]{number of function evaluations}
\acro{rtf}[RTF]{real-time factor}
\acro{mos}[MOS]{mean opinion scores}
\acro{ema}[EMA]{exponential moving average}
\acro{visqol}[ViSQOL]{Virtual Speech Quality Objective Listener}
\acro{rk}[RK]{Runge-Kutta}
\acro{svd}[SVD]{singular value decomposition}
\acro{dnn}[DNN]{deep neural network}
\acro{mse}[MSE]{mean squared error}
\acro{se}[SE]{speech enhancement}
\acro{bwe}[BWE]{bandwidth extension}
\acro{fm}[FM]{flow matching}
\acro{cfm}[CFM]{conditional flow matching}
\acro{jfm}[JFM]{joint flow matching}
\acro{wer}[WER]{word error rate}
\acro{flop}[FLOP]{floating-point operation}
\acro{api}[API]{application programming interface}
\acro{vbdmd}[VB-DMD]{VoiceBank-DEMAND}
\acro{ewv2}[EWv2]{EARS-WHAM v2}
\acro{lrk}[LRK]{Learned Runge-Kutta}
\acro{db}[DB]{Diffusion Buffer}
\acro{ourmethod}[SFM]{Stream.FM}
\end{acronym}

\begin{document}

\title{Real-Time Streamable Generative Speech Restoration with Flow Matching}

\author{
Simon Welker\,{\orcidlink{0000-0002-6349-8462}},
Bunlong Lay\,{\orcidlink{0000-0002-0847-7896}},
Maris Hillemann\,{\orcidlink{0009-0007-5701-8411}},
Tal Peer\,{\orcidlink{0000-0002-8974-9127}},
Timo Gerkmann\,{\orcidlink{0000-0002-8678-4699}},~\IEEEmembership{Senior Member,~IEEE}
\thanks{All authors are with the Signal Processing Group, Dept. of Informatics, Universität Hamburg, 22527 Hamburg, Germany (e-mail: firstname.lastname@uni-hamburg.de).}
}

\maketitle

\begin{abstract}
Diffusion-based generative models have greatly impacted the speech processing field in recent years, exhibiting high speech naturalness and spawning a new research direction. Their application in real-time communication is, however, still lagging behind due to their computation-heavy nature involving multiple calls of large DNNs.

Here, we present Stream.FM, a frame-causal flow-based generative model with an algorithmic latency of 32 milliseconds (ms) and a total latency of 48 ms, paving the way for generative speech processing in real-time communication.
We propose a buffered streaming inference scheme and an optimized DNN architecture, show how learned few-step numerical solvers can boost output quality at a fixed compute budget, explore model weight compression to find favorable points along a compute/quality tradeoff, and contribute a model variant with 24 ms total latency for the speech enhancement task.

Our work looks beyond theoretical latencies, showing that high-quality streaming generative speech processing can be realized on consumer GPUs available today. Stream.FM can solve a variety of speech processing tasks in a streaming fashion: \acl{se}, dereverberation, codec post-filtering, bandwidth extension, STFT phase retrieval, and Mel vocoding.
As we verify through comprehensive evaluations and a MUSHRA listening test, Stream.FM establishes a state-of-the-art for generative streaming speech restoration, exhibits only a reasonable reduction in quality compared to a non-streaming variant, and outperforms our recent work (Diffusion Buffer) on generative streaming \acl{se}
while operating at a lower latency.
\end{abstract}


\section{Introduction}
\label{sec:intro}
Real-time speech processing refers to a type of speech processing that can be performed in an online fashion, where a model produces new parts of the processed output signal as soon as possible after new parts of the input signal arrive. Most applications allow for a fixed latency $\lat$, e.g., few tens of milliseconds (ms) for voice-over-IP (VoIP) communication.
Compared to offline speech processing methods, this fixed latency also requires the system to have a fixed amount of lookahead, i.e., a limited future context. Depending on the task, this may result in a performance degradation compared to an offline method which has potentially all future context available.

While the task of neural \acf{se} has traditionally been considered a predictive task, works from recent years, in particular SGMSE+ \cite{richter2023speech}, have shown that treating \ac{se} as a generative task instead can have several advantages, particularly in perceived speech quality and model generalization.
More broadly, non-additive speech restoration tasks such as \acf{bwe}, dereverberation or \ac{stft} \ac{pr} are naturally well-suited for generative methods, in particular when a significant amount of information is missing from the input signal and must be \emph{regenerated} given only the remaining information \cite{lemercier2023analysing,peer2023diffphase}. Prior work \cite{lemercier2023analysing} empirically supports this idea, showing clear advantages of modern generative methods over predictive methods for such problems. Other tasks which have been successfully tackled with diffusion-based generative methods include neural codec post-filtering \cite{wu2024scoredec,welker2025flowdec}, Mel vocoding \cite{welker2025realtime}, and binaural speech synthesis \cite{liang2025binauralflow}.

However, a key downside to such generative methods is that they are computationally intensive due to the involved numerical solvers evaluating a large \ac{dnn} multiple times. %
This seems to preclude the use of these methods for many real-time scenarios \cite{lay2025diffusion}. However, we show here that real-time and multi-step inference are not necessarily at odds with each other and can be realized on available consumer hardware, given that one uses appropriately buffered inference schemes and matching network architectures.

While many prior works describe their methods as \emph{real-time} \cite{liang2025binauralflow,hsieh2025towards,lu2025realtime}, this is often either \B{(a)} not supported through timings on real hardware, making it unclear whether a streaming implementation is realistically attainable, or \B{(b)} measured based on the \acf{rtf} estimated using offline processing, i.e., using an input utterance duration longer than a single frame, and dividing by the processing time the model takes for the full utterance.
This neglects that, in offline processing, the inference process can make full use of parallelism across the time dimension of the input signal, as well as CUDA kernels which are typically optimized to process single large tensors as quickly as possible rather than many smaller tensors one-by-one.
Such reported offline \acp{rtf} may therefore severely underestimate the actual \ac{rtf} for streaming inference, making it unclear which methods are practically real-time capable in a streaming setting.

With this work, we aim to close this gap and make modern generative streaming-capable methods available to the research community. We propose Stream.FM, a real-time capable streaming generative model based on flow matching \cite{lipman2023flow} that can solve various speech restoration problems with a total latency below 50\,\text{ms}, bringing the high-quality capabilities of diffusion-based generative models to real-time speech processing.
We extend our previous conference publication on real-time streaming Mel vocoding \cite{welker2025realtime}, which \B{(1)} proposes a frame-wise causal DNN and iterative inference scheme for real-time flow matching inference,
also used in this work, 
\B{(2)} realizes streaming inference with a total latency of 48 ms on consumer hardware, and \B{(3)} outperforms established non-causal baselines including HiFi-GAN \cite{kong2020hifigan}.

We expand upon our previous conference publication \cite{welker2025realtime} in the following ways: \B{(4)} we demonstrate wide applicability of our model for general speech restoration with five additional restoration tasks beyond Mel vocoding: speech enhancement, dereverberation, bandwidth extension, codec post-filtering, and STFT phase retrieval; \B{(5)} we newly propose and investigate the use of learned Runge-Kutta \ac{ode} solvers for flow matching to boost output quality under a fixed computational budget; \B{(6)} we provide a comprehensive comparison of attainable streaming latencies of our model and baselines on consumer hardware; \B{(7)} for \ac{se}, we design a joint predictive-generative model inspired by StoRM \cite{lemercier2023storm}; \B{(8)} we explore the use of model weight compression \cite{guo2018decoupling} towards a flexible and favorable choice along the computation/quality tradeoff.%

A full-band variant of the models detailed here, which jointly performs \ac{se} and \ac{bwe}, has been accepted as a Show-and-Tell demo at ICASSP 2026 \cite{welker2026realtimedemo_bwe}. Code\footnote{\url{https://github.com/sp-uhh/streamfm}, published after acceptance.} and audio examples are available online\footnote{\url{https://sp-uhh.github.io/streamfm_examples}}.

\subsection{Related work}\label{sec:related-work}

In 2020, Défossez et al. \cite{defossez2020realtime} proposed a predictive waveform-domain model for causal real-time \ac{se}, which we refer to as \emph{DEMUCS} here. The recent work aTENNuate \cite{pei2025optimized} is a predictive state-space model for real-time \ac{se}. DEMUCS and aTTENuate operate at a similar, slightly larger algorithmic latency than Stream.FM, but no streaming implementation of aTTENuate has been published. 
Dmitrieva and Kaledin \cite{dmitrieva2025hifi} recently proposed HiFi-Stream
for streaming \ac{se}, which uses a block-wise inference scheme at the cost of possible block-edge discontinuities and a higher latency.
MambAttention \cite{kuhne2026mambattention} is a recently proposed model for generalizable speech enhancement, based on a combination of modern state-space models and attention layers, but is not targeted towards streaming inference and uses a non-causal configuration.

Richter et al.~\cite{richter2024causal} first proposed the use of causal \acp{dnn} for diffusion-based \ac{se} and achieved convincing results, but did not target real-time capability on real hardware, and used heavy down- and upsampling along time which leads to large latencies \cite{hedegaard2022cins,lay2025diffusionjpre}.

Lay et al. \cite{lay2025diffusion} proposed the \ac{db}, which realizes a real-time streaming \ac{se} diffusion model on consumer hardware. \ac{db} couples the diffusion time $\dtime$ to physical time $\time$ in a fixed-size frame buffer. %
The buffer is progressively denoised frame-by-frame using one \ac{dnn} call each time, and each time the oldest---then fully denoised---frame in the buffer is output. This realizes a fixed algorithmic latency equal to the total buffer duration, which studies indicate should not be reduced below 180\,ms \cite{lay2025diffusionjpre}. %

Liang et al. \cite{liang2025binauralflow} describe a similar approach to streaming \ac{fm} as detailed here, but only investigate monaural-to-binaural speech upmixing whereas we treat six different speech tasks. The authors determine \acp{rtf} on a 0.683-second snippet which may underestimate streaming \ac{rtf} significantly, see \cref{sec:results:timings-flops-latencies}.
To ensure reproducibility and enable future developments, we provide a more extensive description of the buffered multi-step inference scheme, see \cref{sec:rt-multistep-streaming-diffusion}, as well as a public code repository.

AnyEnhance \cite{zhang2025anyenhance} is a generative speech restoration model using mask-based generative modeling of discrete speech tokens, and addresses various restoration tasks including denoising, dereverberation and bandwidth extension with a single model at a sampling rate of 44.1 kHz. While \cite{zhang2025anyenhance} shows impressive performance, it uses proprietary noise and RIR data which hinders reproducibility, and only a significantly smaller and less powerful model variant\footnote{\url{https://github.com/viewfinder-annn/AnyEnhance-v1}} has been published with code and checkpoints as part of a challenge \cite{zhang2025ccf-sr-challenge}.
AnyEnhance performs enhancement in a highly compressed token space, rather than an uncompressed continuous space like our method. This can aid efficiency, but may also degrade the signal-level similarity to the clean target speech.

A concurrent work \cite{hsieh2025towards} investigates streaming \ac{fm} models for speech restoration, similarly using a U-Net architecture without time-wise downsampling as proposed here. The authors report an algorithmic latency of 20\,ms, but do not provide real-hardware timings, making it unclear whether a streaming model can be practically realized.

Note that all models described above are \B{(a)} not streaming-capable with low latency and/or \B{(b)} have only been demonstrated for speech enhancement -- i.e., background noise removal -- and not for general speech restoration tasks. This shows that a gap exists in the literature regarding streaming-capable models for general speech restoration, which we aim to close with this work.

\section{Background}\label{sec:background}
In this section, we will detail the necessary background and notation for 
this work. We will denote time-domain sequences as lowercase $\auC, \auD \in \mathbb{R}^n$ and their \ac{stft} frame sequences as uppercase $\specC, \specD \in \mathbb{C}^{T \times F}$ with $T$ frames, each having $F$ frequencies. $\auC, \specC$ refer to clean audio and $\auD, \specD$ refer to corrupted audio, $\specE$ indicates an estimate of $\specC$, and $\auE := \iSTFT(\specE)$. We use the indexing notation $\specD[t]$ to indicate the $t$-th frame of $\specD$, and the slicing notation $Y[t_0:t_1]$ to indicate the range of frames in $\specD$ from $t_0$ to $t_1$ (inclusive).

\subsection{Diffusion- and flow-based speech processing}

Diffusion-based speech enhancement was originally introduced in \cite{lu2022conditional,welker2022speech} and first achieved state-of-the-art quality with SGMSE+ \cite{richter2023speech}. Song et al. \cite{song2021scorebased} originally proposed to define \emph{diffusion models} for generative modeling via time-continuous \emph{forward \acp{sde}} that model a continuous mapping from data to noise. Each forward \ac{sde} has a corresponding reverse \ac{sde} resulting directly from mathematical theory \cite{song2021scorebased}, which can be used to map from tractable noise samples to newly generated data. The reverse \ac{sde} involves the \emph{score function} $\nabla_{x} \log(p(x))$ of the data distribution $p(x)$, which is intractable in general but can be learned by a neural network called a \emph{score model} \cite{song2021scorebased}.

For speech enhancement, SGMSE \cite{welker2022speech} and SGMSE+ \cite{richter2023speech} propose specific modified \acp{sde}, modeling speech corruption by interpolating between clean speech $\auC$ and corrupted speech $\auD$ and incrementally adding Gaussian white noise. To produce enhanced speech, one draws a Gaussian white noise sample, adds it to the corrupted speech, and then numerically solves the reverse \ac{sde} starting from this point. We refer the reader to \cite{richter2023speech} for the full detailed description of the training and inference.

Later works introduce \acf{fm} \cite{lipman2023flow}, which is closely related
to diffusion models %
but takes a perspective based on \acp{ode} rather than \acp{sde}. The idea is to learn a model to transport samples from a tractable distribution $q_0(\spec_0)$, e.g., a multivariate Gaussian, to an intractable data distribution $q_1(\spec_1) = \pdata$ by solving the \ac{ode}
\begin{equation}
    \frac{\d}{\d\dtime} \flow(\dtime,\spec) = \vel(\dtime, \flow(\dtime,\spec)) \label{eq:flow-ode}\,,\quad
    \flow(0,\spec) = \spec_0
\end{equation}
starting from a random sample $\spec_0 \sim q_0$. $\flow : [0,1] \times \mathbb{R}^n \to \mathbb{R}^n$ is called the \emph{flow} with the associated \emph{time-dependent vector field} $\vel : [0,1] \times \mathbb{R}^n \to \mathbb{R}^n$, generating a \emph{probability density path} $p_\dtime : \mathbb{R}^n\to\mathbb{R}_{+}$, i.e., a probability density function that depends on an artificial time coordinate $\dtime \in [0,1]$ with $p_0 = q_0$ and $p_1 = q_1$. One can learn a neural network $\lvel$ called a \emph{flow model} to approximate $\vel(\dtime, \cdot)$ with the \emph{conditional flow matching} loss \cite[Eq.~9]{lipman2023flow}:
\begin{equation}\label{eq:flow-matching-objective}
    \loss_{\mathrm{CFM}} = \E_{\spec,\dtime,(\spec_\dtime|\spec)} \left[ \left\lVert \lvel(\dtime,\spec_\dtime) - \vel(\dtime,\spec_\dtime|\spec) \right\rVert_2^2 \right]
\end{equation}
where $\dtime \sim \mathcal{U}(0,1), \spec \sim q_1$ is clean data sampled from a training corpus, $\mathbb{E}_a[b(a)] = \int_{-\infty}^{\infty} b(a) p(a) \d a$ denotes the expectation of $b(a)$ with respect to the distribution $p(a)$ of $a$, and $\theta$ denotes the set of \ac{dnn} parameters. Note that, during training, the expectation is approximated by an empirical average over the training data. The objective \cref{eq:flow-matching-objective}, where $\vel$ is \emph{conditional} on the clean $\spec$, has the same gradients as an intractable objective where $\vel$ is unconditional \cite[Eq.~5]{lipman2023flow}, 
and results in the correct probability path $p_\dtime(\spec_\dtime)$ and flow field $u(\dtime,\spec_\dtime)$ \cite[Sec.~3.1, 3.2]{lipman2023flow}.
FlowDec \cite{welker2025flowdec} showed that, similar to SGMSE \cite{welker2022speech,richter2023speech}, \ac{fm} can be modified to interpolate between the distributions of clean and corrupted data, enabling the use of \ac{fm} for generative signal enhancement. Concretely, we choose the following path, which linearly interpolates from a clean sample $\specC$ to a corrupted sample $\specD$ and adds increasing amounts of Gaussian noise:
\begin{equation}\label{eq:flow-matching-our-prob-path}
    p_\dtime(\spec_\dtime | \specC, \specD)
    := \Gaussian(\spec_\dtime ; (1-\dtime)\specD + \dtime \specC, \Sigma_\dtime)
\end{equation}
where $\Sigma_\dtime = ((1-\dtime) \Sigma_y + \dtime \Sigma_{\text{min}})^2$ and $\Sigma_y, \Sigma_\text{min}$ denote covariance matrices which we assume to be scalar or diagonal. 
We denote a scalar covariance by $\sigma_y^2$, i.e., $\Sigma_y = \sigma_y^2I$ with the identity matrix $I$.
The flow model $\lvel$ can be trained via the \emph{joint flow matching} loss \cite{welker2025flowdec}:
\begin{equation}\label{eq:our-jfm-loss}
\begin{split}
    \loss_{\mathrm{JFM}} = \E_{
        \dtime,(\specC,\specD),(\spec_\dtime|\specC,\specD)
    } \big[
        \big\lVert \lvel(\dtime,\spec_\dtime,\specD) - (\spec_1 - \spec_0) \big\rVert_2^2
    \big]
\end{split}
\end{equation}
where $(\specC,\specD)$ are paired data of clean and corrupted audio sampled from a training corpus. Note that we let each set of $\spec_1, \spec_0$ and $\spec_\dtime$ share a single Gaussian noise sample $\varepsilon \sim \mathcal{N}(0,I)$, as a simple form of training variance reduction through minibatch coupling \cite{pooladian2023multisample}. While \cite{welker2025flowdec} did not use $\Sigma_\text{min}$, we reintroduce it here similar to \cite{lipman2023flow} as we found this to slightly increase training stability.%

The trained network $\lvel$ can then be plugged into the flow \ac{ode} \cref{eq:flow-ode} in place of $\vel$, and this neural \ac{ode} can be solved numerically starting from a sample $\spec_0 \sim p_0$ to perform signal enhancement. This typically requires multiple calls of the network $\lvel$, where the number of calls is referred to as \ac{nfe}.

\subsection{Latency definitions}\label{sec:latency-definitions}
We define the algorithmic latency $\latalg$ as the attainable latency on infinitely fast hardware, and the overall latency of a system as $\lattot = \latalg + \latproc$, where $\latproc$ is the processing time per frame. For a frame-causal \ac{stft}-based method with causal (right-hand) window alignment such as Stream.FM, the algorithmic latency $\latalg$ is the synthesis window length $\winlensynth$ minus one sample divided by the sampling rate $f_s$, i.e., $\latalg = \frac{\winlensynth-1}{f_s}$ \cite{wood2019unsupervised,wang2023verylow}. In typical frame-by-frame processing implementations, $\latproc$ is effectively given exactly via the synthesis frame hop $\hopsynth$, due to the synthesis side waiting for each synthesis frame hop to be complete before producing audio samples. This allows the processing model up to $\latproc \leq \frac{\hopsynth}{f_s}$ of processing time.
We use the same configurations for analysis and synthesis here, hence the analysis window $\winlenana=\winlensynth$ and hop $\hopana=\hopsynth$, so $\lattot = \frac{\winlenana -1 + \hopana}{f_s}$ for our models. We report both $\latalg$ and $\lattot$ for all methods, but note that to attain $\lattot$ on concrete hardware the streaming \ac{rtf} must be reliably below 1, see \cref{sec:results:timings-flops-latencies}.

\begin{figure}[t]
    \centering
    \includegraphics[width=0.8\linewidth]{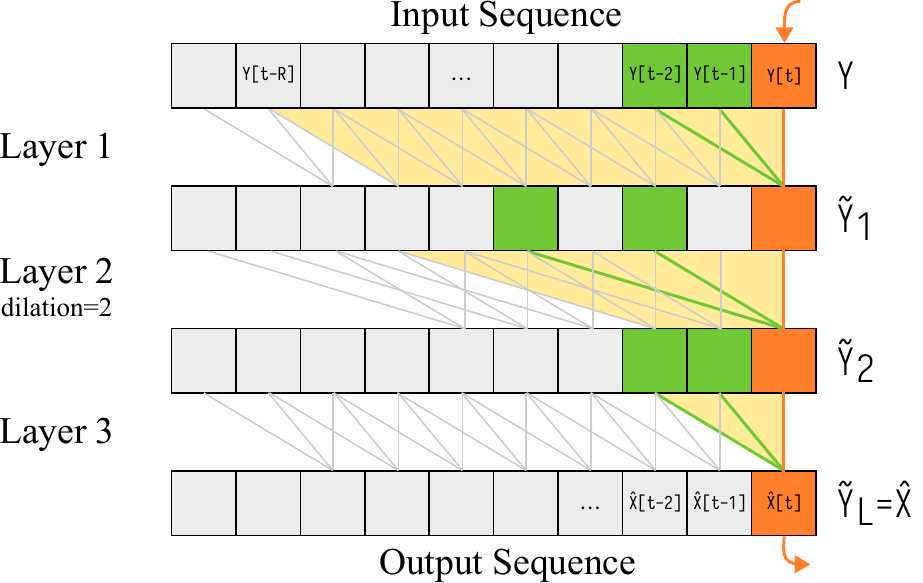}
    \caption{Inference for one new frame (orange) in a simplified frame-causal DNN. While the output frame has a receptive field size (yellow) of 9 in the input, only 3 frames must be evaluated in each layer since all required past results (green) can be stored in a buffer $\mathbf B$.}
    \label{fig:streaming-cnn-concept}
    \vspace{-1.5em}
\end{figure}

\subsection{Explicit Runge-Kutta ODE solvers}\label{sec:background-rk-solvers}
To solve \acp{ode} such as \cref{eq:flow-ode} numerically, a simple method is the Euler method \cite{hairer1993solving}. It starts from $\spec_0$ and $\dtime=0$, discretizes the time interval $\dtime \in [0,1]$ into uniformly spaced points, and iterates the following equation for $N$ iterations until $\tau=1$ using $\Delta\dtime = \frac{1}{N}$:
\begin{equation}\label{eq:euler-step}
    \spec_{\dtime+\Delta\dtime} := \spec_\dtime + \Delta\dtime \cdot \vel(\dtime,\spec_\tau) \approx \spec_\dtime + \Delta\dtime \cdot \lvel(\tau,X_\tau)\,,
\end{equation}
where we inserted the learned flow model $\lvel$ for $\vel$ in the approximate equality. The Euler solver thus requires $\text{NFE}=N$ calls of the \ac{dnn} $\lvel$, but has a relatively large approximation error for a given \ac{nfe} budget compared to more flexible solvers~\cite{hairer1993solving}. %
For high-quality real-time inference, we want our solvers to have a fixed small \ac{nfe} and attain high quality under this budget. To this end, we use explicit Runge-Kutta solvers \cite{hairer1993solving}, which can be parameterized by a
set of coefficients $\mathbf A \in \mathbb{R}^{r \times r}, \mathbf b \in \mathbb{R}^r, \mathbf c \in \mathbb{R}^r$ \cite{butcher1964highorder} where $r$ is the \ac{nfe} per step, and $\mathbf A$ is a strictly lower triangular matrix 
so the solver is explicit
\cite{hairer1993solving}. While these coefficients are usually predefined, in \cref{sec:methods:customrk}, we propose to learn task-optimized $(\mathbf A, \mathbf b, \mathbf c)$ from data.
The iterated equation of such a solver is:
\begin{align}
    G_1 &:= \lvel(\dtime,\spec_\dtime) \label{eq:rk_1} \\
    G_{i>1} &:= \lvel\left(\dtime +c_i \Delta\dtime,\ \spec_\dtime + \Delta\dtime\textstyle\sum_{j=1}^{i-1} a_{ij} G_j \right)\,, \label{eq:rk_k} \\
    \spec_{\dtime+\Delta\dtime} &:= X_\dtime + \Delta\dtime\left(\textstyle\sum_{i=1}^q b_i G_i \right) \label{eq:rk_xtau}
\end{align}
where $a_{ij} = \mathbf A[i,j]$. We set $N=1$ and $\Delta\dtime=1$ so that a single evaluation of \cref{eq:rk_1,eq:rk_k,eq:rk_xtau} is performed, stepping directly to $\dtime=1$. This guarantees $\text{NFE}=r$, with each $G_i$ contributing one evaluation of $\lvel$.

\section{Methods}
\subsection{Multi-step streaming diffusion}
\label{sec:rt-multistep-streaming-diffusion}

When we use a \ac{dnn}
in an inference setting for diffusion- or flow-based signal enhancement, processing a noisy frame sequence $\specD$ into a clean estimate $\specE$, we use \ac{ode} or \ac{sde} solvers. For \ac{fm}, \ac{ode} solvers are used. Within these solvers, we call the learned \ac{dnn}, $\lvel$, $N$ times in sequence, starting from a noisy sequence $\specD_0 = \specD+\varepsilon$ with some independent noise sample $\varepsilon$. We assume here that $\lvel$ has a finite receptive field of size $R$ and is frame-causal, i.e., there exists no $t$ such that the frame $\specE[t]$ depends on any input frame $Y[t+n], ~ n>0$.
For ease of illustration, we use the equidistant Euler solver \cref{eq:euler-step} for ODEs with $N$ steps in the following and assume the model $\lvel$ was trained with the \ac{jfm} objective \cref{eq:our-jfm-loss}. This inference process generates intermediate partially-denoised sequences $Y_1, Y_2, \ldots, Y_{N-1} \in \mathbb{R}^{T \times F}$ and finally the fully denoised sequence $Y_N$, which each depend on their respective prior sequence as:
\begin{align}
    \specD_1[t] &= \specD_0[t] + \Delta\tau \cdot \lvel(0, \specD_0[t-R:t]), \label{eq:diff-solver-1}\\
    \specD_2[t] &= \specD_1[t] + \Delta\tau \cdot \lvel(\Delta\tau, \specD_1[t-R:t]),\quad\ldots, \label{eq:diff-solver-2}\\
    \specD_N[t]   &= \specD_{N-1}[t] + \Delta\tau \cdot  \lvel((N-1)\Delta\tau, \specD_{N-1}[t-R:t]), \label{eq:diff-solver-n}
\end{align}
where %
the first parameter $\tau \in [0,1]$ to $\lvel(\tau, \cdot)$ is the continuous diffusion time and $\Delta\tau = \frac{1}{N}$ is the discretized diffusion timestep. After completing this process, $\specD_N$ is fully denoised and we can treat it as the clean sequence estimate $\specE:=\specD_N$. Since $\specD_0$ is just the addition of $\specD$ and Gaussian noise $\varepsilon$, recursively collapsed, this implies the following effective dependency of $\specE$ on $\specD$:
\begin{equation}
    \specE[t] = V_\theta(\specD[t - N \cdot R:t]),
\end{equation}
where $V_\theta$ represents the evaluation of the whole procedure in \cref{eq:diff-solver-1,eq:diff-solver-2,eq:diff-solver-n}. %
Given this, it may seem that real-time streaming diffusion inference with a large DNN $\lvel$ is not viable: to generate a single output frame $\specE[t]$ we must run the entire diffusion process backwards, doing so we must process all $NR$ frames, and we cannot exploit time-wise parallelism for efficiency since frames come in one-by-one.

One approach to circumvent this is the Diffusion Buffer \cite{lay2025diffusion}, see \cref{sec:related-work}. In a follow-up preprint \cite{lay2025diffusionjpre}, the authors show that with a different training loss and direct-prediction inference, frames can be output earlier to reduce latency. However, this approximates the reverse process that the \ac{dnn} was trained for, with the approximation quality decreasing as one decreases the latency. This results in significant quality degradations when targeting the same algorithmic latency as our method, see \cref{tab:streamflow-se} ($d=0$), suggesting the need for a different approach in low-latency settings.

\subsection{Efficient real-time streaming diffusion inference}\label{sec:efficient-rt-diffusion-inference}
We provide a scheme that can perform streaming frame-wise inference while incurring no algorithmic latency and avoiding any redundant computations. For this purpose, we assume that $\lvel$ has a specific form, namely that of a causal convolutional neural network containing $L$ stacked causal convolution layers $\mathcal{C}_\lay$ each with stride 1, dilation $d_\lay$, and a kernel $K_\lay \in \mathbb{R}^{c_{o,\lay} \times c_{i,\lay} \times k_\lay}$ of size $k_\lay$ with input channels $c_{i,\lay}$ and output channels $c_{o,\lay}$.
The \ac{dnn} may also contain any operations that are point-wise in time, e.g., nonlinearities and specific types of normalization layers. Each such layer has receptive field size $R_\lay = (k_\lay -1)d_\lay + 1$, and the overall receptive field size of $\lvel$ is $R = 1 +\sum_{\lay=1}^{L}R_\lay-1$.
The first key observation here is that a causal convolution of kernel size $k$ and dilation $d$ depends on its input sequence only at $k$ time points in $\specD[t-(k-1)d:t]$, and immediately when its inputs are available, its output is fixed. Even when stacking multiple convolutions, the past outputs of $\mathcal{C}_{\lay-1}$ never need to be recomputed. 
This lets us perform local caching throughout the DNN, where every layer $\mathcal{C}_{\lay}$ keeps an internal rolling buffer $B_\lay \in \mathbb{R}^{c_{i,\lay} \times (R_\lay-1)}$ which contains only $R_\lay-1$ past input frames, with $R_\lay-1 \ll R$.

When we receive a new input frame $\specD[t]$ to the DNN after previously having seen the input frames $\specD[0],\ldots,\specD[t-1]$, we only need to evaluate each layer's convolution kernel $K_\lay$ at the single latest time index $t$ on the buffer $B_\lay$ concatenated with the newest frame, producing a single output frame $\tilde{\specD}_\lay$ per layer, see \cref{fig:streaming-cnn-concept}:
\begin{align}
    \tilde{\specD}_{1}[t] &= \varphi_1(K_1 \star \specD[
    \underbrace{t-R_1:t-1}_{=B_1}, t]), \\
    \tilde{\specD}_{2}[t] &= \varphi_2(K_2 \star \tilde{\specD}_{1}[\underbrace{t-R_2:t-1}_{=B_2}, t]),\quad\ldots, \\
    \tilde{\specD}_{L}[t] &= \varphi_L(K_L \star \tilde{\specD}_{L-1}[\underbrace{t-R_L:t-1}_{=B_L}, t]),
\end{align}
where $\star$ is the (possibly dilated) convolution evaluated only for a single output frame, yielding an output in $\mathbb{R}^{c_{o,\lay} \times 1}$, and each $\varphi_\lay$ represents arbitrary extra operations between convolutions assumed to be point-wise in time.
We can then set $\tilde{\specD}_{L}[t]$ as the output frame of this single DNN call. Note that all $k_\lay, d_\lay$ can be chosen arbitrarily and independently without affecting the efficiency of this scheme.
This idea has also been explored for real-time video processing
\cite{hedegaard2022cins}. The authors note that strided convolutions incur a delay to their respective downstream module and hence incur algorithmic latency, so we avoid their use here and set all time-wise convolution strides to 1.

The second key observation is that this scheme can be straightforwardly extended to diffusion/flow model inference with multiple DNN calls by tracking $N$ independent collections of cache buffers, one for each DNN call, resulting in $(N \cdot L)$ cache buffers in total. We denote this collection of buffers as $\mathbf B$ where $\mathbf B_{n,\lay} \in \mathbb{R}^{(R_l-1) \times F}$, $n=1,\ldots,N$ and $\lay=1,\ldots,L$. Starting from the initial value $\specD_0[t]$, we can then process the following sequences $\tilde{Y}_{n,l}$ for each $n$-th model call and $l$-th layer:
\begin{gather*}
    \left.
    \begin{aligned}
        \tilde{\specD}_{1,1}[t] &= \varphi_1(K_1 \star [\mathbf B_{1,1}, \specD_0[t]]),\quad\ldots, \\
        \specD_1[t] := \tilde{\specD}_{1,L}[t] &= \varphi_L(K_L \star [\mathbf B_{1,L}, \tilde{\specD}_{1,L-1}[t]]),\\
    \end{aligned}
    \right\} \text{ Call 1 }\\
    \left.
    \begin{aligned}
        \tilde{\specD}_{2,1}[t] &= \varphi_1(K_1 \star [\mathbf B_{2,1}, \specD_1[t]]),\quad\ldots, \\
        \specD_2[t] := \tilde{\specD}_{2,L}[t] &= \varphi_L(K_L \star [\mathbf B_{2,L}, \tilde{\specD}_{2,L-1}[t]]),
    \end{aligned}
    \right\} \text{ Call 2 }\\
    \scalebox{0.6}{\vdots}\hspace{8em}\notag\\
    \specD_N[t] := \tilde{\specD}_{N,L}[t] = \varphi_L(K_L \star [\mathbf B_{N,L}, \tilde{\specD}_{N,L-1}[t]]),\hspace{2.5em}
\end{gather*}
where $[\cdot, \cdot]$ denotes concatenation along the time dimension. For notational simplicity, we conceptually include the ODE solver step, see \cref{eq:diff-solver-1,eq:diff-solver-2,eq:diff-solver-n}, in each $\varphi_L$. After each operation, the respective buffer $\mathbf B_{n,\lay}$ is shifted to drop the oldest frame and the new frame is inserted on the right, see \cref{fig:streaming-cnn-concept}. The buffers are all initialized with zeros, matching the training if all convolution layers use zero-padding.

This scheme now performs the same computations as an offline implementation, only spread across physical time. Importantly, it yields exactly the same result---up to tiny numerical floating-point errors---as feeding the entire sequence to the causal model at once in an offline, batch-wise fashion. This fact enables the use of standard batched training and the subsequent approximation-free use of the trained weights for streaming inference, in contrast to, e.g., the Diffusion Buffer when outputting frames early for lower latency \cite{lay2025diffusionjpre}.

We note that this scheme can also be used with \acp{sde} with one minor modification. For \ac{sde} solvers, a new noise sample $\varepsilon_n$ is drawn at every solver step $n=1,\ldots,N$. By redefining
     $\specD_n[t] := \tilde{\specD}_{n,L}[t] + \tilde{\sigma}_n\varepsilon_n[t]$,
where $\tilde{\sigma}_n$ is the level of noise added according to the \ac{sde} solver and diffusion schedule, and treating this as another point-wise operation in time, no other changes to the inference are required.

\subsection{DNN architecture}\label{sec:dnn-modifications}
We design a custom frame-causal CNN without strided convolutions, as a modified variant of the NCSN++ architecture \cite{song2021scorebased}. NCSN++ is a non-causal 2D U-Net architecture with multiple residual blocks per level, a progressive down/upsampling path, as well as non-causal attention layers.
For a detailed description and illustration of the specific NCSN++ configuration we build upon, we refer the reader to SGMSE+ \cite{richter2023speech}. As in \cite{richter2023speech}, our DNN receives the audio signal as a complex-valued STFT (a 2D time-frequency signal), with the real and imaginary parts mapped to two real-valued \ac{dnn} input/output channels.
We perform the following modifications and simplifications:
\begin{enumerate}
    \item We perform down- and upsampling only along frequency, never along time. To increase time context, we replace time-strided with time-dilated causal convolutions (dilation of 2).
    \item We keep the FIR filters for anti-aliased down- and upsampling \cite{song2021scorebased} along frequency, but remove the FIR filtering along time.
    \item We remove all attention layers. It is in principle possible to include causal attention layers while staying streaming-capable by using causal attention with a fixed-size window for constant runtime and memory complexity, but for simplicity, our main models do not use attention. We briefly investigate adding such a layer in our ablation study, see \cref{sec:ablation-study}.
    \item We replace the original GroupNorm \cite{wu2018group} with a custom sub-band grouped BatchNorm (\emph{SGBatchNorm}), inspired by \cite{chang2021subspectral}. SGBatchNorm
    performs joint grouping along channels and frequencies and normalizes each group. We use four frequency groups, and the channel grouping of NCSN++ \cite{song2021scorebased}.
    Prior work \cite{richter2024causal} used time-cumulative group normalization, which determines statistics frame-by-frame as a time-varying IIR filter. In contrast, SGBatchNorm determines running-average batch statistics during training which are frozen and thus time-invariant during inference.
    \item In contrast to \cite{richter2024causal}, we keep the progressive input down/upsampling paths \cite{song2021scorebased}, by simply using causal convolutions and the SGBatchNorm described above.
    \item We remove an extra residual block on each level on the upward path, which was used to process a skip connection from the \emph{input} of each corresponding block on the downward path. We only keep the skip connection from the \emph{output} of the corresponding downward block.
    \item To fuse skip connections, we always employ addition, while NCSN++ \cite{song2021scorebased} used addition on the downward and channel concatenation on the upward path. Our proposed addition significantly reduces the number of features on the upward path.
    \item We do not use an exponential moving average (EMA) of the weights, since we found our models to perform well without it.
\end{enumerate}
All modifications described above are implemented in the DNN architecture we make available in our public codebase. Further details on engineering-level optimizations of our models towards real-time inference can be found in \crefnobox{sec:model-impl-and-optimizations} of the supplementary material.

\subsection{Predictive-generative speech enhancement}\label{sec:pred-gen-se}
Inspired by Lemercier et al.~\cite{lemercier2023storm}, for our \ac{se} task we use a joint predictive-generative approach, where an \emph{initial predictor} network $D_\eta$ with parameters $\eta$ is first trained to map $\specD$ to an initial estimate $Z := D_\phi(y)$, and $Z$ is then used instead of $\specD$ for defining and training the generative model. For $D_\eta$, we use the same \ac{dnn} architecture as for the \ac{fm} model, but remove all diffusion time conditioning layers. To train $D_\eta$, we use the following loss:
\begin{align}\label{eq:loss-initial-predictor}
    \loss_\text{pred}(\auIE, \auC) &:= \frac{1}{2}\lVert z - \auC \rVert_1 +
        \frac{1}{2} \loss_\text{MR-STFT}(\auIE, \auC),\\
    \loss_\text{MR-STFT}(\auIE, \auC) &:= \sum_{w=1}^{N_w} \big\lVert \ |\STFT_w(\auIE)| - |\STFT_w(\auC)| \ \big\rVert_1,
\end{align}
where $\auIE := \iSTFT(\specIE)$ and $\loss_\text{MR-STFT}$ is a multi-resolution magnitude \ac{stft} $L_1$ loss similar to \cite{yamamoto2020parallelwavegan}, using a set of $\STFT_w$ with different windows $w$ with $N_w$ window configurations. We use $N_w=4$ Hann windows with $\winlenana \in \{256, 512, 768, 1024\}$ and 50\% overlap.

\subsection{Custom learned low-NFE ODE solvers}\label{sec:methods:customrk}
Given a fixed budget of DNN evaluations per frame (\ac{nfe}), we develop specialized \ac{lrk} ODE solvers (see \cref{sec:background-rk-solvers}) to optimize the achievable quality without model retraining or finetuning, at virtually no increase of computations during inference.
Conceptually, our proposal is related to \cite{guo2022personalized} from numerical literature, but differs in the following ways: we use loss functions specific to speech processing instead of a simple \ac{mse}; we set $\Delta\tau=1$ to guarantee a fixed low \ac{nfe}; since intermediate results are not valid clean speech estimates, we only calculate the loss on the single endpoint $\hat{x}_1 = \hat{s}$; our integrated function $\lvel$ is high-dimensional and given by a neural network, rather than a simple function from low-dimensional numerical \ac{ode} problems.
We train the scheme's parameters $\{\mathbf A, \mathbf b, \mathbf c\}$, given a pretrained flow matching model $v_{\theta}$ with frozen weights $\theta$, by solving the ODE \eqref{eq:flow-ode} with the current \ac{rk} scheme and treating the final output as the clean speech estimate via $\auE := \iSTFT(\spec_1)$. To optimize $\{\mathbf A, \mathbf b, \mathbf c\}$, we backpropagate through the whole solved ODE path and multiple \ac{dnn} calls to determine the gradients, using the following loss:
\begin{equation}
\begin{aligned}\label{eq:custom-rk-loss}
    \mathcal{L}_{\text{RK}} := -\text{SBS}(\auE, \auC) + 0.001 \cdot \text{MR-LS-MSE}(\auE, \auC)
\end{aligned}
\end{equation}
where SBS refers to SpeechBERTScore \cite{saeki24speechbertscore} and MR-LS-MSE refers to a multi-resolution log-magnitude STFT \ac{mse} loss similar to \cite{yamamoto2020parallelwavegan}. The motivation for using negative SpeechBERTScore \cite{saeki24speechbertscore} is to reduce phonetic hallucinations that generative methods can suffer from \cite{richter2024causal,deoliveira2026gibberish}, and the motivation for MR-LS-MSE is to increase fine high-frequency detail, which \ac{fm} models tend to lose in low-\ac{nfe} settings \cite{welker2025flowdec}. %
For this loss, we use window sizes $\winlenana \in \{320, 512, 640\}$ and 75\% overlap.
We further consider a loss oriented towards signal fidelity, replacing the SpeechBERTScore term with a differentiable\footnote{\url{https://github.com/audiolabs/torch-pesq}} PESQ loss \cite{martin2018deep}:
\begin{equation}
\begin{aligned}\label{eq:custom-rk-loss-pesq}
    \mathcal{L}_{\text{RK,PESQ}} := \mathcal{L}_{\mathrm{torchPESQ}}(\auE, \auC) + 0.001 \cdot \text{MR-LS-MSE}(\auE, \auC)
\end{aligned}
\end{equation}

We ensure by construction of the optimized parameters that $\sum_j a_{ij} = c_i, \sum_i b_i = 1$, and $0.05 \leq b_i \leq 1$ for all $1 \leq i,j \leq r$ to ensure that all model calls contribute to the final estimate.
The constraint $\sum_i b_i = 1$ guarantees the consistency of our schemes, as well as convergence with order of (at least) 1 \cite{butcher2016numericalode}.
The region of absolute stability of each \ac{lrk} scheme can be found in \crefnobox{fig:lrk-absolute-stability-regions} of the supplementary material.
Our \ac{lrk} schemes do not guarantee a convergence order larger than 1 since the learning-based procedure of determining the coefficients makes it unlikely that they will satisfy the required algebraic conditions \cite[Eq.~(2.21)]{hairer1993solving} for order-2 and above, which we argue is acceptable since our solvers do not need to be general-purpose. We find that they nonetheless transfer reasonably well, see \cref{sec:lrk-solvers-analysis}.
We clip $c_i \leq 0.85$ to avoid evaluating the \ac{fm} model close to $\tau\approx1$ where the training target \eqref{eq:our-jfm-loss} is unstable, and put a quadratic penalty loss on each $a_{ij}$ outside the range $[-2,2]$.
We train $\{\mathbf A, \mathbf b, \mathbf c\}$ with the Adam optimizer at a learning rate of $10^{-3}$, a batch size of 10, and 2-second audio snippets, for 25,000 steps on our training dataset.
For the 4 \ac{nfe} available under our runtime budget in the \ac{se} task, we initialize $\{\mathbf{A}, \mathbf{b}, \mathbf c\}$ from Kutta's four-stage 3/8 scheme \cite{hairer1993solving}. For all other tasks, we have five \ac{nfe} available due to the lack of a predictive DNN,
and we use one Ralston-2 step followed by one Ralston-3 step \cite{ralston1962runge} for initialization. 
We list the \ac{rk} parameters learned for each task in \crefnobox{sec:learned-rk-coefs} of the supplementary material.

\subsection{Model compression through weight decoupling}\label{sec:model-compression}
We make use of DNN compression methods to reduce the computational costs of each network call. Using this, we aim to improve the quality given a fixed computational budget, or to decrease the number of computations at some reasonable decrease in output quality.
There are various methods to this end including weight pruning, quantization, or model distillation, but here we specifically follow the decoupling approach of Guo et al. \cite{guo2018decoupling} to approximate the large 2D convolution weight tensors $\mathbf{W} \in \mathbb{R}^{n_o \times n_i \times k_h \times k_w}$ with output/input channels $n_o$, $n_i$ and kernel size $k_h \times k_w$, since such convolutions incur most of the computational effort in our networks.

The authors first show that a 2D convolution can be losslessly decomposed into one depthwise and one pointwise convolution without increasing the algorithmic complexity. They further show a direct connection between these two layers and the \acf{svd} of the weights for each input channel, $\mathbf W_{:,i} \in \mathbb{R}^{n_o \times k_h \times k_w}$, reshaped to a matrix $\widetilde{\mathbf{W}_{:,i}} \in \mathbb{R}^{n_o \times (k_h k_w)} = USV^\top$. $U$ and $SV^\top$ are determined via the \ac{svd} and map directly to the weight tensors for the depthwise and the pointwise convolution, respectively.
This leads to a \ac{svd}-based compression of pretrained convolution layers, by truncating the SVD to rank $J \leq K = \min(n_o,k_h k_w)$ where $J=K$ indicates no compression.
The truncated matrices also map directly to the weights of smaller depthwise and pointwise convolution layers.

We apply this weight compression to all $3\times3$ Conv2d layers with $n_o \geq 9$ in our network.
As also proposed in \cite{guo2018decoupling} we then fine-tune the compressed models for 25,000 training steps. We deviate slightly from the authors' representation, choosing $U\sqrt{S}$ for the depthwise and $\sqrt{S}V^\top$ for the pointwise weights to spread the singular values across the two layers and improve fine-tuning stability.

\section{Experiments}\label{sec:experiments}
\subsection{Speech restoration tasks}
We provide here a description of the six speech restoration tasks we investigate. See \cref{tab:corruption-models} for the full list of corruption and feature representation operators we use in practice.
\begin{table}
    \centering
    \caption{Corruption and feature representation operators for our speech restoration tasks. $*$ indicates convolution, $(\cdot\downsample f)$ and $(\cdot\upsample f)$ indicate down/upsampling by a factor $f$, respectively, $\text{Dec}$/$\text{Enc}$ refer to the encoder/decoder of an audio codec, and $M$ is a per-frame STFT$\to$Mel matrix with $M^\dagger$ as its Moore-Penrose pseudoinverse. $+0j$ indicates embedding of real numbers into the complex plane.}
    \label{tab:corruption-models}
    \begin{tabular}{lll}
        \toprule
         \B{Task} & \B{Problem} & \B{Corruption model} \\
         \midrule
         1 & Speech Enhancement & $Y = \STFT\{x+n\}$ \\
         2 & Dereverberation & $Y = \STFT\{x * h\}$ \\
         3 & Codec Post-Filtering & $Y = \STFT\{\text{Dec}(\text{Enc}(x))\}$ \\
         4 & Bandwidth Extension & $Y = \STFT\{(x \downsample f) \upsample f\}$ \\
         5 & STFT Phase Retrieval & $Y = \big|\STFT\{x\}\big| + 0j$ \\
         6 & Mel Vocoding & $Y = \big|M^\dagger \left(M \big| \STFT\{x\}\big|\right)\big| + 0j$ \\
         \bottomrule
         \vspace{-1.5em}
    \end{tabular}
\end{table}

\subsubsection{Speech Enhancement}\label{sec:experiments:se}
For our \acf{se} task, the signal corruption model is $\auD = \auC + n$, where $\auC$ is the clean audio and $n$ is some uncorrelated background noise.
As the dataset, we use \acl{ewv2}\footnote{\label{earsv2}see \url{https://github.com/sp-uhh/ears_benchmark} for the v2 release.} (\acs{ewv2}) \cite{richter2024ears}, downsampled to 16\,kHz. We also use the \ac{ewv2} clean utterances as the dataset for all following tasks except dereverberation.

\subsubsection{Dereverberation}
We investigate speech dereverberation using the EARS-Reverb v2 dataset \cite{richter2024ears}. The signal corruption model is $\auD = \auC * h$, where $*$ indicates time-domain convolution and $h \in \mathbb{R}^{l_h}$ is a sampled \ac{rir}.

\subsubsection{Codec Post-Filtering}
Inspired by ScoreDec \cite{wu2024scoredec} and FlowDec \cite{welker2025flowdec}, we investigate the use of Stream.FM as a post-filter for a low-bitrate speech codec.
FlowDec \cite{welker2025flowdec} introduces \ac{fm}-based generative post-filtering, proposing a non-adversarially trained variant of the neural codec DAC \cite{kumar2023dac}. %
Since DAC is non-causal and computationally expensive, we use the Lyra V2 codec\footnote{\url{https://github.com/google/lyra}} 
instead, which is built for streaming speech coding on consumer devices.
Lyra V2 produces one frame every 20\,ms at our chosen bitrate of 3.2 kbit/s. To align the codec frames with our model, we change the \ac{stft} parameters to use 40\,ms windows and 20\,ms hops.

\subsubsection{Bandwidth Extension}
We train a model to perform \ac{bwe} from speech downsampled to sampling frequencies of 8\,kHz and 4\,kHz, leading to a frequency cutoff at 4\,kHz and 2\,kHz, respectively. To generate $\auD$, we downsample each $\auC$ randomly to either 8 or 4\,kHz.

\subsubsection{STFT Phase Retrieval}\label{sec:experiments:stft-pr}
Peer et al. have shown with DiffPhase \cite{peer2023diffphase} that SGMSE+ \cite{richter2023speech} can be modified to solve an \ac{stft} \ac{pr} task, resulting in very high reconstruction quality. We extend this idea to our streaming setting, using only 50\% \ac{stft} overlap instead of the 75\% overlap used in DiffPhase. We compare our method against the non-causal DiffPhase \cite{peer2023diffphase}, which we reconfigure with $\winlensynth = 510, \hopsynth = 255$ for exactly 50\% overlap. We use these \ac{stft} parameters for compatibility with the DiffPhase code and DNN. We retrain DiffPhase using the same data, batchsize, and number of optimizer steps as for our method. We further compare against a family of streaming \ac{stft} \ac{pr} algorithms proposed by Peer et al. \cite{peer2024flexible}. 

\subsubsection{Mel Vocoding}
As we show in \cite{welker2025realtime}, the ideas of DiffPhase \cite{peer2023diffphase} can be extended to streaming Mel vocoding through a small change in the corruption model. Instead of treating the phaseless STFT magnitudes $|X|$ as the corrupted signal \cite{peer2023diffphase}, we additionally subject them to a lossy Mel compression as follows:
\begin{equation}
    X_{\text{mel}}[t] = \big|M^\dagger \left(M \big| X[t]\big|\right)\big| + 0j
\end{equation}
where $M \in \mathbb{R}^{F_{\text{mel}} \times F_{\text{STFT}}}$ is the Mel matrix mapping STFT frames to Mel frames, $M^\dagger$ is its Moore-Penrose pseudoinverse, and $X[t]$ denotes the single magnitude spectrogram frame at frame index $t$.
We follow the Mel configuration of HiFi-GAN \cite{kong2020hifigan} in the 16\,kHz variant from SpeechBrain \cite{ravanelli2024speechbrain}, but to reduce the latency, we use 32\,ms windows instead of 64\,ms while keeping the 16\,ms hop length.

\subsection{Data and data representation}
We use the EARS dataset \cite{richter2024ears} as the basis of all our problem variants and model trainings, resampled to a sampling frequency of $f_s=$\,16\,kHz.
In evaluations for \ac{se}, we also use the \ac{vbdmd} \cite{valentini2016investigating} dataset, also resampled to 16 kHz.
Unless otherwise noted, we use an \ac{stft} with a 512-point periodic $\sqrt{\text{Hann}}$-window (32 ms), a 256-point hop length (16 ms, 50\% overlap), and magnitude compression with exponent $\alpha = 0.5$ as in \cite{welker2022speech,richter2023speech}. Different from these works, we use an orthonormal STFT and do not apply an additional scaling. Since a 512-point window leads to 257 frequency bins, for ease of \ac{dnn} processing, we discard the Nyquist band to retrieve frames with 256 frequency bins, and pad it back with zeros before applying the inverse \ac{stft}.%

For \ac{se}, similar to \cite{richter2024causal}, during training we peak-normalize $\auC$ and $\auD$ independently and apply a random negative gain between -12 and 0\,dB to $\auD$, so that the model learns to perform automatic gain control.
For all other tasks, we normalize $\auC$ and $\auD$ jointly based on the peak magnitude of $\auD$, assuming that a roughly constant input-output level relationship is available in these tasks.
As the \ac{fm} process hyperparameter $\Sigma_y$ \eqref{eq:flow-matching-our-prob-path}, we empirically set scalar $\sigma_y=0.05$ for \ac{se}, $\sigma_y=0.25$ for \ac{stft} \ac{pr} and Mel vocoding, and $\sigma_y=0.35$ for dereverberation and codec post-filtering. For \ac{bwe}, we follow \cite{welker2025flowdec} and determine a heuristic per-frequency-band diagonal covariance matrix $\Sigma_y$ to avoid adding noise in the preserved low-frequency bands and to allow easier regeneration of the low-energy high-frequency bands. We set $\sigma_{\min}=0.001$ for all tasks except \ac{bwe} where $\Sigma_{\min} = 0.001\cdot\Sigma_y$.

\subsection{DNN configuration and training}\label{sec:dnn-configuration}
For Stream.FM, we parameterize our architecture described in \cref{sec:dnn-modifications} with two residual blocks per level and four U-Net levels with [128, 256, 256, 256] channels, respectively, leading to 27.9\,M parameters when used as an \ac{fm} backbone \ac{dnn}.
For the initial predictor network $D_\phi$ in the \ac{se} task, we remove all time-conditioning layers and reduce the complex input channels from two to one, resulting in 24.6\,M parameters. For the non-causal \ac{fm} baseline models, we use the original NCSN++ architecture \cite{song2021scorebased}, here also parameterized with four U-Net levels in the same channel configuration and also two residual blocks per level (38.7\,M parameters).

For each task, we train a flow model $\lvel$ using \cref{eq:our-jfm-loss} for 150,000 steps on two NVIDIA RTX A6000 GPUs, using 2-second random snippets with a batch size of 12 per GPU. We use the SOAP optimizer \cite{vyas2025soap} which was recently found to perform well for diffusion model training \cite{schaipp2025optimization}.%
We use a cosine annealing learning rate schedule with a maximum learning rate of $\lambda=5\times10^{-4}$ and linear warmup for the first 1,000 steps, clamping the scheduled $\lambda$ to a minimum value of $10^{-6}$. 
We use gradient clipping with a maximum norm of $\lVert\nabla\rVert_{\mathrm{max}}=3$ for all tasks except for \ac{se} with $\lVert\nabla\rVert_{\mathrm{max}}=1$ and codec post-filtering with $\lVert\nabla\rVert_{\mathrm{max}}=5$, based on empirical gradient norm inspection.

To train the \ac{se} model, we first train the initial predictor $D_\phi$ for 150,000 steps using \cref{eq:loss-initial-predictor} with the SOAP optimizer \cite{vyas2025soap} at a constant learning rate of $\lambda = 3\times10^{-3}$. We then freeze the initial predictor during \ac{fm} model training. For \ac{se}, we also train a lower-latency joint predictive-generative Stream.FM model, reconfiguring the \ac{stft} with 16\,ms frames and 8\,ms hops for a total latency of only 24\,ms, keeping all other settings the same.

\subsection{Evaluation}

\subsubsection{Baseline methods}\label{sec:eval:baselines}
As streaming-capable baseline methods for \ac{se}, we use DEMUCS \cite{defossez2020realtime}, DeepFilterNet3 \cite{schroeter2023deepfilternet3}, HiFi-Stream \cite{dmitrieva2025hifi}, CleanUMamba \cite{groot2025cleanumamba}, Diffusion Buffer \cite{lay2025diffusionjpre}, and a SEMamba \cite{semamba} variant modified for streaming and trained \cite{lay2025diffusionjpre} on \ac{ewv2} \cite{richter2024ears}.
We further include the aTENNuate \cite{pei2025optimized} method for which no streaming model variant has been published, hence we only evaluate its offline variant. As offline-only baselines, we use the non-causal FM model described in \cref{sec:dnn-configuration},
SBVE \cite{jukic2024schrodinger,richter2025investigating}, MambAttention \cite{kuhne2026mambattention}, and AnyEnhance \cite{zhang2025anyenhance,zhang2025ccf-sr-challenge}.
We use the official streaming implementations for DEMUCS\footnote{\url{https://github.com/facebookresearch/denoiser}, \texttt{master64} checkpoint.}, HiFi-Stream\footnote{\url{https://github.com/KVDmitrieva/source_sep_hifi}, \texttt{hifi\textunderscore{}fms} checkpoint.}, DeepFilterNet3\footnote{\url{https://github.com/Rikorose/DeepFilterNet}, \texttt{DeepFilterNet3} checkpoint.}, and CleanUMamba\footnote{\url{https://github.com/lab-emi/CleanUMamba}, \texttt{E8\textunderscore{}pruned-5M} checkpoint.}, and the official Python package for aTENNuate\footnote{\url{https://pypi.org/project/attenuate/}}. For SBVE, we use the checkpoint\footnote{\url{https://github.com/sp-uhh/sgmse}} trained on 16 kHz EARS-WHAM \cite{richter2024ears} and \ac{vbdmd} \cite{valentini2016investigating}.%

As baselines for other tasks, we use: AnyEnhance \cite{zhang2025anyenhance,zhang2025ccf-sr-challenge} for dereverberation, \ac{bwe} and codec artifact removal; for dereverberation, a 48 kHz SGMSE+ model \cite{richter2024ears} trained on EARS-Reverb v1; for \ac{stft} \ac{pr}, DiffPhase \cite{peer2023diffphase}; for Mel vocoding, HiFi-GAN \cite{kong2020hifigan} with the 16 kHz SpeechBrain checkpoint \cite{ravanelli2024speechbrain}.

\subsubsection{Latency determination}\label{sec:eval:latency-check}
Theoretical derivations of latencies can be misleading in complex systems. We thus determine $\latalg$ in an end-to-end fashion:
At every single index in a 2-second example input waveform, we set the value (only at this index) to IEEE 754 \nan{} (not a number), %
and let each model produce an enhanced output waveform. \nan{}s are infectious, i.e., any operation involving a \nan{} produces a \nan{} output, a fact we use to detect dependencies of every output sample on each input sample. We sweep across all input indices and determine the maximum index difference between the affected input index and the first index for which the output is also \nan{}, giving us $\latalg$. We report $\latalg=\infty$ if we find that the latency grows arbitrarily large with the input waveform length.

\subsubsection{Metric evaluation}\label{sec:eval:metrics}
As intrusive metrics, we report wideband PESQ \cite{Rix2001PESQ}, ESTOI \cite{jensen2016estoi}, SI-SDR \cite{roux2019sdr} and \ac{lsd}\footnote{as in \url{https://github.com/haoheliu/ssr_eval}, 32\,ms Hann window, 75\% overlap.} \cite{liu2022neural}. We further report the \ac{wer} using the QuartzNet15x5Base-En model \cite{kriman2020quartznet} from the NeMo toolkit \cite{kuchaiev2019nemo} as the speech recognition backend, using the model's transcripts of the clean audios as the reference. As non-intrusive metrics, we report NISQA \cite{mittag2021nisqa}, WVMOS \cite{andreev2023hifipp}, and DistillMOS \cite{stahl2025distillmos} which we refer to as \emph{DiMOS} for brevity.
We evaluate all model outputs at 16 kHz, downsampling to 16 kHz as necessary, e.g., for AnyEnhance \cite{zhang2025anyenhance}.

\subsubsection{Listening experiments}
We conduct two MUSHRA-like listening experiments \cite{mushra2014method}, one for \ac{se} and one for \ac{bwe}, each with 12 participants who gave informed consent. We asked participants to rate the overall quality (0--100) of 8 randomly sampled utterances, as reconstructed by each method.
For \ac{se}, we compare predictive-generative \acf{ourmethod} against non-streaming \ac{fm}, both with 4 Euler steps, Diffusion Buffer \cite{lay2025diffusionjpre} at $d \in \{0,9\}$, and DEMUCS \cite{defossez2020realtime}.
For \ac{bwe}, we compare \ac{ourmethod} against non-streaming \ac{fm}, both with 5 Euler steps, against \ac{ourmethod} with a learned RK5 solver. We use the noisy/downsampled utterance as the low anchor for \ac{se}/\ac{bwe}, respectively.%

\subsubsection{Runtime performance evaluation}\label{sec:eval:runtime-performance}
We perform all model runtime evaluations using a single laptop with an \emph{NVIDIA RTX 4080 Laptop} GPU. We measure the number of \acp{flop} using the PyTorch \texttt{torch.utils.flop{\textunderscore}counter} module, and the wall clock timings using \texttt{torch.cuda.Event}.

\section{Results and discussion}\label{sec:results}
\begin{figure}
    \centering
    \begin{subfigure}{0.55\linewidth}
        \centering
        \includegraphics[scale=0.57]{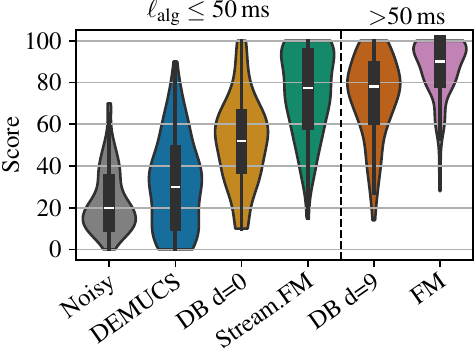}
        \caption{Speech enhancement}
        \label{fig:listening-results-se}
    \end{subfigure}%
    \hfill
    \begin{subfigure}{0.45\linewidth}
        \centering
        \includegraphics[scale=0.57]{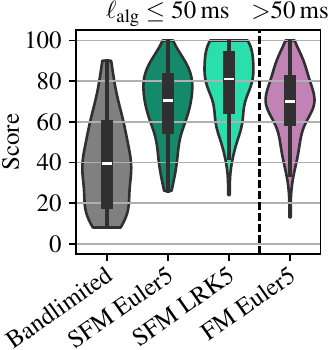}
        \caption{Bandwidth extension}
        \label{fig:listening-results-bwe}
    \end{subfigure}
    \caption{Violin plots of the scores listeners assigned to examples from each method in the listening experiments for \B{(a)} speech enhancement and \B{(b)} bandwidth extension. \emph{SFM} is Stream.FM, \emph{FM} is the flow matching baseline, and \emph{DB} is the Diffusion Buffer \cite{lay2025diffusionjpre}. Note that FM has $\latalg=\infty$ and DB with $d=9$ has $\latalg\approx180$\,ms.
    \vspace{-1.5em}%
    }
    \label{fig:listening-results-all}
\end{figure}

\begin{table*}
    \centering
    \caption{%
    Mean metrics for speech enhancement on \ac{ewv2} (16 kHz) \cite{richter2024ears}.
    Our main model (SFM) is evaluated using different ODE solvers (\emph{Euler1}, \emph{Euler4}, etc.) with the number indicating the number of solver steps, and compared against several streaming and non-streaming baselines.
    \enquote{LRK4} refers to a four-stage learned Runge-Kutta solver, see \cref{sec:methods:customrk}.
    In the NFE column, \emph{1+n} indicates a single call for the initial predictor and $n$ calls for the flow model.
    Methods marked with $^*$ use author-provided model checkpoints trained on different data.
    Best within a group \B{bold}, second best \U{underlined}, worse than input in \R{red}.
    $\latalg$ and $\lattot$ in milliseconds.
    }
    \label{tab:streamflow-se}
    \begin{tabular}{lrrrrrrrrrrrr}
        \toprule
        \B{Method} & \B{NFE} & \B{PESQ} & \B{ESTOI} & \B{SI-SDR} & \B{DiMOS} & \B{WVMOS} & \B{NISQA} & \B{WER$\downarrow$} & \B{LSD$\downarrow$} & $\latalg$ & $\lattot$ & \B{Params} \\
        \midrule
        Noisy & - & 1.24 & 0.64 & 5.4 & 2.58 & 1.20 & 1.95 & 32.8\% & 2.24 & - & - \\
        \midrule

        \multicolumn{8}{l}{\scriptsize \textsc{Stream.FM} (32 ms)} \\
        SFM Euler1 & 1+1 & 2.18 & \B{0.84} & \B{15.2} & \B{3.91} & 2.65 & \U{4.43} & \U{19.5\%} & 1.39 & 32 & 48 & 52.5M \\ %
        SFM Euler4 & 1+4 & 2.09 & \U{0.83} & 14.3 & \U{3.88} & 2.72 & \B{4.50} & 21.8\% & 1.29 & 32 & 48 & 52.5M \\  %
        SFM Midpoint2 & 1+4 & 2.02 & 0.82 & 13.3 & 3.66 & 2.72 & 4.28 & 23.6\% & 1.38 & 32 & 48 & 52.5M \\  %
        SFM LRK4 \eqref{eq:custom-rk-loss} & 1+4 & 2.24 & 0.82 & 14.0 & 3.81 & \B{3.00} & 4.06 & 20.2\% & 1.42 & 32 & 48 & 52.5M \\  %
        SFM LRK4 \eqref{eq:custom-rk-loss-pesq} & 1+4 & \U{2.30} & \U{0.83} & 14.1 & 3.70 & 2.79 & 4.04 & 20.1\% & 1.36 & 32 & 48 & 52.5M \\
        Initial Predictor $D_\phi$ & 1 & 2.11 & 0.79 & 13.5 & 3.64 & 2.60 & 3.56 & 25.4\% & 1.42 & 32 & 48 & 24.6M \\
        
        \arrayrulecolor{black!20}\midrule\arrayrulecolor{black}

        \rowcolor{white}
        \multicolumn{8}{l}{\scriptsize \textsc{Streaming Baselines}} \\
        Diffusion Buffer $d=0$ \cite{lay2025diffusionjpre} & 1 & 1.75 & 0.74 & 10.9 & 2.75 & 2.17 & 2.45 & 27.3\% & 1.45 & 32 & 48 & 22.2M \\ %
        Diffusion Buffer $d=9$ \cite{lay2025diffusionjpre} & 1 & 2.09 & 0.81 & \U{15.0} & 3.66 & 2.56 & 3.81 & 21.7\% & 1.57 & 176 & 192 & 22.2M \\ %
        SEMamba (causal) \cite{semamba,lay2025diffusionjpre} & 1 & \B{2.61} & 0.82 & 8.8 & 3.75 & 2.65 & 3.25 & \B{18.6\%} & \U{1.16} & 25 & 31 &  1.24M \\  %
        DeepFilterNet3$^*$ \cite{schroeter2023deepfilternet3} & 1 & 1.76 & 0.70 & 8.8 & 3.29 & 2.81 & 3.37 & \R{37.0\%} & \R{2.46} & 40 & 50 & 2.14M \\  %
        DEMUCS$^*$ \cite{defossez2020realtime} & 1 & 1.95 & 0.79 & 13.0 & 3.46 & \B{3.00} & 2.61 & 27.1\% & 1.55 & 41 & 57 & 33.5M \\  %
        CleanUMamba$^*$ \cite{groot2025cleanumamba} & 1 & 1.98 & 0.79 & 13.5 & 3.49 & \U{2.84} & 2.72 & 27.5\% & \B{1.08} & 48 & 64 & 4.9M \\  %
        HiFi-Stream$^*$ \cite{dmitrieva2025hifi} & 1 & \R{1.21} & \R{0.37} & \R{-3.8} & \R{1.94} & 1.49 & \R{1.50} & \R{69.7\%} & 1.57 & 256 & 384 & 1.6M \\   %

        \midrule
        \rowcolor{white}
        \multicolumn{8}{l}{\scriptsize \textsc{Lower-Latency Stream.FM (16 ms)}} \\
        SFM 16ms Euler1 & 1+1 & \U{2.07} & \B{0.83} & \B{14.8} & \U{3.72} & 2.59 & \U{4.43} & \B{20.5\%} & \U{1.42} & 16 & 24 & 52.5M \\   %
        SFM 16ms Euler4 & 1+4 & 2.00 & 0.81 & 14.3 & 3.64 & 2.63 & 4.44 & 22.7\% & \B{1.28} & 16 & 24 & 52.5M \\
        SFM 16ms LRK4 \eqref{eq:custom-rk-loss-pesq} & 1+4 & \B{2.16} & 0.81 & 13.7 & 3.47 & \B{2.73} & 3.94 & \U{21.3\%} & \B{1.28} & 16 & 24 & 52.5M \\

        \midrule
        \rowcolor{white}
        \multicolumn{8}{l}{\scriptsize \textsc{Non-Streaming Methods}} \\
        FM Euler1 & 1+1 & \U{2.36} & \B{0.86} & \B{16.8} & 4.20 & 2.73 & 4.37 & \B{16.4\%} & 1.53 & $\infty$ & $\infty$ & 73.7M \\
        FM Euler4 & 1+4 & \B{2.41} & \B{0.86} & 16.1 & \U{4.34} & 2.82 & \B{4.50} & \U{18.4\%} & \U{1.31} & $\infty$ & $\infty$ & 73.7M \\
        SBVE \cite{jukic2024schrodinger,richter2025investigating} & 60 & 2.15 & 0.84 & 12.9 & \B{4.39} & \B{3.14} & 4.18 & \U{18.4\%} & 1.60 & $\infty$ & $\infty$ & 65.6M \\
        AnyEnhance$^*$ (44.1 kHz) \cite{zhang2025anyenhance} & 1 & 1.80 & 0.72 & 6.0 & 4.02 & 2.63 & 3.81 & 30.7\% & 1.41 & $\infty$ & $\infty$ & 45.7M \\
        MambAttention$^*$ \cite{kuhne2026mambattention} & 1 & 2.06 & 0.73 & 7.4 & 3.62 & 2.71 & 3.24 & 31.3\% & \B{1.17} & $\infty$ & $\infty$ & 2.33M \\
        aTENNuate$^*$ \cite{pei2025optimized} & 1 & 1.86 & 0.72 & 9.0 & 2.95 & \U{2.86} & 2.37 & \R{33.1\%} & 1.42 & $\infty$ & $\infty$ & 0.8M \\  %
        \bottomrule
    \end{tabular}
\end{table*}
\rowcolors{1}{white}{white}

\begin{table}
    \centering
    \caption{\acl{vbdmd} speech enhancement benchmark (16 kHz) \cite{valentini2016investigating}.
    \enquote{DB} refers to the Diffusion Buffer \cite{lay2025diffusionjpre}.
    DB, SFM, FM, and Causal SEMamba were trained on \ac{ewv2}.
    All models configured as in \cref{tab:streamflow-se}.
    $\latalg$ in milliseconds.
    }
    \label{tab:vb-dmd-results}
    \resizebox{\linewidth}{!}{
    \begin{tabular}{lrrrrr}
        \toprule
        \B{Method} & \B{PESQ} & \B{ESTOI} & \B{SI-SDR} & \B{DiMOS} & \B{$\latalg$} \\
        \midrule
        Noisy & 1.97 & 0.79 & 8.4 & 2.58 & - \\
        SFM Euler4 & \U{2.72} & \B{0.85} & 13.4 & \B{3.88} & 32 \\
        SFM LRK4 \eqref{eq:custom-rk-loss} & 2.69 & \U{0.84} & 13.3 & \U{3.81} & 32 \\
        SFM LRK4 \eqref{eq:custom-rk-loss-pesq} & \U{2.72} & 0.84 & 13.0 & 3.70 & 32 \\

        \arrayrulecolor{black!20}\midrule\arrayrulecolor{black}
        
        DB $d=0$ \cite{lay2025diffusionjpre} & 2.42 & 0.81 & 13.1 & 2.75 & 32 \\
        DB $d=9$ \cite{lay2025diffusionjpre} & 2.45 & \U{0.84} & 14.5 & 3.66 & 176 \\
        SEMamba \cite{semamba,lay2025diffusionjpre} & \B{3.18} & 0.79 & \R{5.8} & 3.52 & 31 \\

        DeepFilterNet3 \cite{schroeter2023deepfilternet3} & 2.71 & \U{0.84} & \B{17.3} & 3.34 & 40 \\  %
        DEMUCS \cite{defossez2020realtime} & 2.60 & \B{0.85} & 15.1 & 3.46 & 41 \\  %
        CleanUMamba \cite{groot2025cleanumamba} & 2.64 & \B{0.85} & \U{16.4} & 3.47 & 48 \\
        HiFi-Stream \cite{dmitrieva2025hifi} & 2.48 & 0.83 & 14.1 & 3.35 & 256 \\  %

        \midrule
        FM Euler4 & 2.86 & \U{0.86} & 14.1 & \B{4.34} & $\infty$ \\
        SBVE \cite{jukic2024schrodinger,richter2025investigating} & \B{3.06} & \B{0.89} & \B{19.2} & \U{3.90} & $\infty$ \\
        AnyEnhance \cite{zhang2025ccf-sr-challenge,zhang2025anyenhance} & 2.72 & 0.81 & 10.8 & 3.84 & $\infty$ \\
        aTENNuate \cite{pei2025optimized} & \U{2.97} & 0.84 & \U{16.2} & 2.95 & $\infty$ \\  %
        \bottomrule
    \end{tabular}
    }
    \vspace{-2em}
\end{table}

\subsection{Speech enhancement}\label{sec:results:se}
We show the metric results for the \ac{se} task in \cref{tab:streamflow-se}.
We find that all \ac{ourmethod} variants attain the best or second-best values among streaming-capable methods in almost all metrics and exhibit particularly strong improvements in non-intrusive DistillMOS, WVMOS and NISQA. We also list the metrics for the outputs of the initial predictor $D_\phi$, which demonstrate that the subsequent flow matching stage with the LRK4 \eqref{eq:custom-rk-loss-pesq} solver clearly improves all metrics.
When the \acf{db} baseline \cite{lay2025diffusionjpre} is configured to use the same low algorithmic latency ($d=0$) as \ac{ourmethod}, where \ac{db} uses only a single diffusion step per frame, \ac{ourmethod} with four Euler steps shows strong advantages over \ac{db}. \ac{ourmethod} also performs similar to or better than the higher-latency \ac{db} variant $(d=9)$ with 10 diffusion steps, the main model proposed in \cite{lay2025diffusionjpre}.

Causal SEMamba \cite{semamba,lay2025diffusionjpre} achieves the best PESQ and WER values but falls short on SI-SDR and non-intrusive metrics, suggesting excellent denoising ability but suboptimal speech quality. CleanUMamba \cite{groot2025cleanumamba}, while not trained on \ac{ewv2}, shows the best overall LSD value and decent performance in most other metrics except NISQA. HiFi-Stream \cite{dmitrieva2025hifi} and aTTENuate \cite{pei2025optimized} do not achieve clear improvements over the noisy mixtures and in particular worsen the \ac{wer}. This may indicate an overfitting to their respective training datasets, which can be seen as reasonable under their small \ac{dnn} parameter budgets. The non-causal generative method AnyEnhance \cite{ahn2025fastenhancer} has good non-intrusive metric values but falls short on all intrusive metrics, which may be related to performing enhancement in a compressed token space. 
SBVE \cite{jukic2024schrodinger,richter2025investigating}, also a non-causal generative method, shows good quality overall but is mostly outperformed by our non-causal FM model, and uses an expensive 60 NFE.
In \crefnobox{fig:snr-split-evaluation} of the supplementary material, we further provide a detailed metric evaluation figure split per input SNR, which reveals that all methods exhibit relatively stable behavior over the full input SNR range.

Comparing different ODE solvers for Stream.FM, we see that a single Euler step (\emph{Euler1}) shows excellent performance in this task, but 4 Euler steps (\emph{Euler4}) slightly improve most non-intrusive metrics and \ac{lsd}, at some decrease in intrusive metric scores.
The \ac{lrk} solver (\emph{LRK4}) with loss \eqref{eq:custom-rk-loss} strongly improves PESQ, WVMOS and WER over Euler4, but does not yield a metric improvement across the board. Using the PESQ-based loss \eqref{eq:custom-rk-loss-pesq} instead to train the solver improves PESQ, WER, and LSD, though at the cost of some decrease in non-intrusive metrics.
The non-streaming \ac{fm} baseline outperforms \ac{ourmethod}, but the quality degradation of the streaming models is expected due to the small look-ahead, and is on acceptable levels.
Our $\latalg=16$\,ms lower-latency \ac{ourmethod} variant shows relatively minor quality reduction compared to the main \ac{ourmethod} model, proving the viability of our method for \ac{se} in even lower-latency settings ($\lattot \approx 24$\,ms).

In \cref{tab:vb-dmd-results}, we show metrics on the classic \ac{vbdmd} benchmark \cite{valentini2016investigating}. \ac{ourmethod} exhibits the generalization capabilities expected of diffusion-based \ac{se} models \cite{richter2023speech} and attains the best ESTOI and DistillMOS among streaming methods. \ac{ourmethod} is second-best in PESQ, outperformed only by causal SEMamba \cite{semamba} which however degrades SI-SDR below the noisy input and does not improve ESTOI.
Both \ac{lrk} solver variants, which were trained on \ac{ewv2} data, exhibit reasonable cross-dataset transfer and perform similarly to Euler4.

In the listening experiment results, depicted in \cref{fig:listening-results-se}, \ac{ourmethod} is clearly preferred over all low-latency baselines and has similar median ratings as the higher-latency \acf{db} method ($d=9,\ \latalg\approx180$\,ms), with a slightly higher upward and downward spread of scores. %
The non-streaming \ac{fm} baseline receives an excellent median score around 90. %
DEMUCS \cite{defossez2020realtime} is not rated well in comparison, possibly indicating a failure to generalize to different data.

\subsection{Dereverberation}\label{sec:results:derev}
\begin{table*}
    \caption{Metric evaluations for the restoration tasks of dereverberation, codec post-filtering, and bandwidth extension.
    Best within a group \B{bold}, second best \U{underlined}, worse than input in \R{red}.
    $\latalg$ and $\lattot$ in milliseconds.
    }
    \begin{subtable}[t]{\textwidth}
    \centering
    \caption{Dereverberation task (EARS-Reverb v2 test set).
    SGMSE+ \cite{richter2023speech} was evaluated with the official checkpoint trained on EARS-Reverb\textunderscore{}v1 \cite{richter2024ears} in 48 kHz, and AnyEnhance \cite{zhang2025anyenhance} uses the official checkpoint with inputs resampled to 44.1 kHz. Both baselines were evaluated using outputs resampled to 16 kHz.
    }
    \label{tab:streamflow-derev}
    \begin{tabular}{lrrrrrrrrrrrr}
        \toprule
        \B{Method} & \B{NFE} & \B{PESQ} & \B{ESTOI} & \B{SI-SDR} & \B{DiMOS} & \B{WVMOS} & \B{NISQA} & \B{WER$\downarrow$} & \B{LSD$\downarrow$} & $\latalg$ & $\lattot$ & \B{Params} \\
        \midrule
        Reverberant       & - & 1.32 & 0.58 & -16.6 & 3.02 & 2.02 & 2.11 & 20.1\% & 1.21 & - & - & - \\
        SFM Euler1 & 1 & 1.63 & 0.73 & -14.2 & 3.29 & 2.17 & 3.03 & \R{23.6\%} & \R{1.49} & 32 & 48 & 27.9M \\
        SFM Euler5 & 5 & \U{2.01} & \B{0.79} & \B{-13.3} & \B{3.76} & \B{2.49} & 3.59 & 16.8\% & 1.12 & 32 & 48 & 27.9M \\
        SFM Midpoint2 & 4 & 1.94 & 0.78 & -14.3 & \U{3.70} & 2.35& \U{3.62} & 17.3\%  & \B{0.98} & 32 & 48 & 27.9M \\
        SFM LRK5 \eqref{eq:custom-rk-loss} & 5 & 1.91 & 0.78 & \U{-13.5} & 3.49 & 2.44 & 3.29 & \U{16.6\%} & \U{1.01} & 32 & 48 & 27.9M \\
        SFM LRK5 \eqref{eq:custom-rk-loss-pesq} & 5 & \B{2.05} & \B{0.79} & \U{-13.5} & 3.68 & \U{2.48} & \B{3.67} & \B{15.9\%} & 1.05 & 32 & 48 & 27.9M \\
        \midrule
        FM Euler5 & 5 & \B{2.31} & \B{0.85} & \B{-11.7} & \B{3.77} & \B{2.43} & \U{3.47} & \B{11.4\%} & \B{1.01} & $\infty$ & $\infty$ & 38.7M \\
        SGMSE+ (48 kHz) \cite{richter2024ears} & 60 & \U{1.95} & \U{0.77} & \U{-15.5} & \U{3.47} & \U{2.37} & \B{3.53} & \U{15.2\%} & \U{1.11} & $\infty$ & $\infty$ & 64.7M \\
        AnyEnhance$^*$ (44.1 kHz) \cite{zhang2025ccf-sr-challenge,zhang2025anyenhance} & 1 & 1.53 & 0.65 & \R{-17.1} & 3.15 & 2.10 & 2.84 & \R{27.1\%} & \R{1.46} & $\infty$ & $\infty$ & 45.7M \\
        \bottomrule
    \end{tabular}
    \end{subtable}
    \begin{subtable}[t]{\textwidth}
    \centering
    \vspace{1em}
    \caption{Lyra V2 codec post-filtering task on \ac{ewv2} test set clean utterances.
    }
    \label{tab:streamflow-lyra}
    \begin{tabular}{lrrrrrrrrrrrr}
        \toprule
        \B{Method} & \B{NFE} & \B{PESQ} & \B{ESTOI} & \B{SI-SDR} & \B{DiMOS} & \B{WVMOS} & \B{NISQA} & \B{WER$\downarrow$} & \B{LSD$\downarrow$} & $\latalg$ & $\lattot$ & \B{Params} \\
        \midrule
        Decoded & - & 2.00 & 0.76 & 1.6 & 3.08 & 2.60 & 2.68 & \B{13.6\%} & \B{1.01} & - & - & - \\
        SFM Euler1 & 1 & \R{1.80} & \R{0.63} & \B{4.9} & \R{2.18} & \R{1.81} & \R{2.45} & \R{44.3\%} & \R{5.84} & 40 & 60 & 27.9M \\
        SFM Euler5 & 5 & \U{2.55} & \B{0.80} & 3.1 & \U{4.00} & \U{2.93} & \U{3.96} & \R{16.3\%} & \R{1.49} & 40 & 60 & 27.9M \\
        SFM Midpoint2 & 4 & 2.38 & \U{0.79} & 1.9 & \B{4.09} & 2.80 & \B{4.11} & \R{\U{16.0\%}} & \R{1.25} & 40 & 60 & 27.9M \\
        SFM LRK5 \eqref{eq:custom-rk-loss} & 5 & 2.27 & 0.77 & 1.8 & 3.94 & 2.71 & 3.90 & \R{16.3\%} & \R{\U{1.09}} & 40 & 60 & 27.9M \\
        SFM LRK5 \eqref{eq:custom-rk-loss-pesq} & 5 & \B{2.58} & \B{0.80} & \U{3.9} & 3.83 & \B{2.97} & 3.74 & \R{17.0\%} & \R{1.68} & 40 & 60 & 27.9M \\
        \midrule
        FM Euler5 & 5 & \B{2.56} & \B{0.80} & \B{3.0} & \B{4.14} & \B{2.92} & \B{3.97} & \R{16.1\%} & \R{1.41} & $\infty$ & $\infty$ & 38.7M \\
        AnyEnhance$^*$ (16 kHz) \cite{zhang2025anyenhance,zhang2025ccf-sr-challenge} & 1 & \R{1.90} & \R{0.74} & \R{0.5} & 3.76 & 2.62 & 3.53 & \R{18.4\%} & \R{1.26} & $\infty$ & $\infty$ & 45.7M \\
        \bottomrule
    \end{tabular}
    \end{subtable}
    \begin{subtable}[t]{\textwidth}
    \centering
    \vspace{1em}
    \caption{Bandwidth extension task ($\{2, 4\}$ kHz $\to 8$ kHz of frequency content) on \ac{ewv2} test set clean utterances.
    }
    \label{tab:streamflow-bwe}
    \begin{tabular}{lrrrrrrrrrrrr}
        \toprule
        \B{Method} & \B{NFE} & \B{PESQ} & \B{ESTOI} & \B{SI-SDR} & \B{DiMOS} & \B{WVMOS} & \B{NISQA} & \B{WER$\downarrow$} & \B{LSD$\downarrow$} & $\latalg$ & $\lattot$ & \B{Params} \\
        \midrule
        Bandlimited & - & \U{3.51} & 0.84 & 15.9 & 3.09 & 2.21 & 2.93 & 19.4\% & 2.24 & - & - & - \\
        SFM Euler1 & 1 & 3.22 & \U{0.92} & \B{16.8} & 3.43 & 2.41 & 3.32 & 15.3\% & 1.95 & 32 & 48 & 27.9M \\
        SFM Euler5 & 5 & 3.37 & \B{0.94} & \U{16.5} & 4.07 & 2.96 & 3.76 & 12.3\% & 1.26 & 32 & 48 & 27.9M \\
        SFM Midpoint2 & 4 & 3.10 & \B{0.94} & 16.0 & \U{4.18} & \U{3.01} & \U{3.97} & \U{12.0\%} & \U{1.10} & 32 & 48 & 27.9M \\
        SFM LRK5 \eqref{eq:custom-rk-loss} & 5 & 3.02 & \B{0.94} & 15.3 & \B{4.19} & \B{3.02} & \B{3.99} & \B{10.5\%} & \B{1.00} & 32 & 48 & 27.9M \\
        SFM LRK5 \eqref{eq:custom-rk-loss-pesq} & 5 & \B{3.71} & \B{0.94} & \B{16.8} & 3.89 & 2.81 & 3.46 & 11.5\% & 1.45 & 32 & 48 & 27.9M \\

        \midrule
        FM Euler5 & 5 & \B{3.52} & \B{0.94} & \B{16.3} & 4.28 & 3.00 & 3.77 & \B{11.8\%} & \B{1.29} & $\infty$ & $\infty$ & 38.7M \\
        AnyEnhance$^*$ (44.1 kHz) \cite{zhang2025ccf-sr-challenge,zhang2025anyenhance} & 1 & \R{2.49} & 0.85 & \R{6.9} & \B{4.33} & \B{3.19} & \B{4.13} & 18.1\% & 1.37 & $\infty$ & $\infty$ & 45.7M \\
        \bottomrule
        \vspace{-1.5em}
    \end{tabular}
    \end{subtable}
\end{table*}

For dereverberation, the results are shown in \cref{tab:streamflow-derev}. \ac{ourmethod} achieves a consistent improvement over the reverberant audio for all solvers except Euler1, showing that inference with $N>1$ is useful in this non-additive restoration problem. While the \ac{lrk} solver trained with \eqref{eq:custom-rk-loss} does not yield a clear improvement, the \ac{lrk} solver with a PESQ-based loss \eqref{eq:custom-rk-loss-pesq} leads to the best results in PESQ, NISQA, and WER, and otherwise performs similarly to Euler5. The non-causal \ac{fm} baseline shows a clear advantage, particularly in PESQ and WER, which we argue is expectable since future information is useful to estimate \ac{rir} characteristics and suppress reverberation, see also our ablation study in \cref{sec:ablation-study}.
The non-causal FM baseline outperforms both AnyEnhance \cite{zhang2025anyenhance} and SGMSE+ \cite{richter2023speech,richter2024ears} in all metrics except NISQA, where it is close to SGMSE+ (difference of 0.06), and SFM (Euler5) also performs slightly better than SGMSE+ while spending 12$\times$ fewer NFE and being streaming-capable.
Note that these two baselines were provided with full-band speech. This may be considered a more difficult task, but the full-band inputs may also contain additional helpful information for faithfully reconstructing the low-frequency bands, and we discard all additionally estimated high-frequency information for evaluation.

\subsection{Codec post-filtering}\label{sec:results:lyra}
For codec post-filtering, we list the results in \cref{tab:streamflow-lyra}. \ac{ourmethod} can greatly improve both intrusive and non-intrusive metrics over the plain Lyra V2 decoder, at the cost of a small increase in \ac{wer}. The learned Runge-Kutta scheme fails to yield an improvement here except in \ac{lsd}.
We conjecture that the loss \eqref{eq:custom-rk-loss} is suboptimal here since the Lyra decoder outputs already have good phoneme and frequency envelope preservation. Using the PESQ-based loss \eqref{eq:custom-rk-loss-pesq} instead improves intrusive metrics (PESQ, ESTOI, SI-SDR) and WVMOS slightly over Euler5, but clearly degrades all other metrics. We suspect that this task is especially difficult to optimize further due to the complicated ambiguities induced by the heavy compression of the Lyra V2 codec.
Notably, the non-streaming \ac{fm} baseline shows only marginal quality improvements over \ac{ourmethod} at five Euler steps here, which may be related to the streaming nature of the Lyra V2 codec itself.
AnyEnhance \cite{zhang2025anyenhance}, which was also trained for codec artifact removal but only for MP3-encoded speech, degrades all intrusive metrics and WER below the decoded audio, but improves all non-intrusive metrics and, from informal listening, does somewhat improve the perceived audio quality.

\subsection{Bandwidth extension}\label{sec:results:bwe}
We list the bandwidth extension metrics in \cref{tab:streamflow-bwe}. The Euler1 solver is now clearly the worst option, with a significant gap in non-intrusive metrics and WER compared to higher-\ac{nfe} solvers.
The AnyEnhance baseline \cite{zhang2025anyenhance} attains very good non-intrusive metric results, but clearly falls short of both our SFM and our FM model in all intrusive metrics (PESQ, ESTOI, SI-SDR, LSD) and WER. Its SI-SDR value of 6.9 is particularly low, which may be related to AnyEnhance performing enhancement in a compressed token space.
Notably, the learned RK5 solver performs best, with a clear improvement in all metrics except PESQ and SI-SDR. LRK5 based on \eqref{eq:custom-rk-loss} also clearly improves WER, suggesting that it reconstructs the semantic content better. This confirms the usefulness of our proposal in this highly ambiguous restoration problem. The improvements from this learned solver are also supported by the listening experiment \cref{fig:listening-results-bwe}, where it receives the highest scores.
Note that PESQ may be misleading here, as the bandlimited audios receive the best PESQ scores, except for LRK5 with the PESQ-based loss \eqref{eq:custom-rk-loss-pesq}. While this LRK5 variant achieves the best PESQ (3.71), it is worse than Euler5 in all non-intrusive metrics and LSD. This is reflected in the outputs of this solver lacking high-frequency detail, see our audio example website. Thus, the solver learned with the PESQ-based objective does not adequately solve the desired bandwidth extension task.

\subsection{STFT phase retrieval / Mel vocoding}\label{sec:results:stftpr-and-mel}
For \ac{stft} \ac{pr} and Mel vocoding, see \cref{tab:streamflow-diffphase,tab:streamflow-melvocoder}, which are both highly ill-posed non-linear inverse problems, the behaviors of the solvers are similar as for \ac{bwe}, but the differences are more pronounced. Euler1 produces unusable estimates here, with worse metrics even than naive zero-phase estimates.
The learned solvers show a clear advantage, particularly in WER and LSD.
On \ac{stft} \acl{pr}, both Stream.FM and the non-causal FM baseline perform better than the non-causal DiffPhase \cite{peer2023diffphase}, except in \ac{wer} and \ac{lsd}, while using fewer parameters and 6 times fewer \ac{nfe}. 
On Mel vocoding, our streaming models outperform HiFi-GAN on all metrics except LSD, see also our conference publication \cite{welker2025realtime}.
Overall, Stream.FM shows excellent performance in these two tasks, with PESQ ($>4.1$) and ESTOI ($\geq0.96$) values approaching the optimum.

\begin{table*}
    \centering
    \caption{Metric evaluation for STFT phase retrieval and Mel vocoding on \ac{ewv2} test set clean utterances.
        RTISI-DM refers to \cite{peer2024flexible} using the Difference Map algorithm \cite{elser2003phaseretrieval} with hyperparameter $\beta = 1.75$, which we grid-searched as the optimum for 50 iterations. RTISI has no parameters.
        DiffPhase \cite{peer2023diffphase} uses matched training, see \cref{sec:experiments:stft-pr}.
        For RTISI and RTISI-DM, \emph{NFE} refers to the number of algorithm iterations.
        Best within a group \B{bold}, second best \U{underlined}, worse than input in \R{red}.
        $\latalg$ and $\lattot$ in milliseconds.
    }
    \begin{subtable}[t]{\textwidth}
    \centering
    \caption{%
        STFT phase retrieval task with 50\% overlap, and no lookahead for all methods except non-causal FM.
    }
    \label{tab:streamflow-diffphase}
    \begin{tabular}{lrrrrrrrrrrrr}
        \toprule
        \B{Method} & \B{NFE} & \B{PESQ} & \B{ESTOI} & \B{SI-SDR} & \B{DiMOS} & \B{WVMOS} & \B{NISQA} & \B{WER$\downarrow$} & \B{LSD$\downarrow$} & $\latalg$ & $\lattot$ & \B{Params} \\
        \midrule
        Zero-phase & - & 1.31 & 0.68 & -34.6 & 1.73 & 1.86 & 1.42 & 14.9\% & 1.19 & - & - & - \\
        RTISI \cite{beauregard2005efficient} & 50 & 3.08 & 0.90 & -28.3 & 3.28 & 2.76 & 3.03 & 5.2\% & 0.71 & 32 & 48 & 0 \\
        RTISI-DM \cite{peer2024flexible} & 50 & 3.35 & 0.91 & -27.0 & 3.86 & 2.83 & 3.56 & 4.9\% & 0.73 & 32 & 48 & 1 \\
        SFM Euler1 & 1 & 1.58 & \R{0.58} & \B{1.7} & 1.90 & \R{1.70} & 2.05 & \R{61.8\%} & \R{6.72} & 32 & 48 & 27.9M \\
        SFM Euler5 & 5 & \B{4.24} & \B{0.97} & \U{-1.7} & \U{4.30} & \U{3.26} & 4.13 & \U{3.7\%} & 0.76 & 32 & 48 & 27.9M \\
        SFM Midpoint2 & 4 & 4.05 & \U{0.96} & -2.5 & \U{4.30} & 3.15 & \B{4.15} & \U{3.7\%} & \U{0.68} & 32 & 48 & 27.9M \\
        SFM LRK5 \eqref{eq:custom-rk-loss} & 5 & \U{4.22} & \B{0.97} & -2.3 & \B{4.40} & \B{3.27} & 4.11 & \B{3.0\%} & \B{0.61} & 32 & 48 & 27.9M \\
        \midrule
        FM Euler5 & 5 & \B{4.38} & \B{0.98} & \B{-1.2} & \B{4.37} & \B{3.25} & \B{4.10} & \U{2.9\%} & \U{0.65} & $\infty$ & $\infty$ & 38.7M \\
        DiffPhase \cite{peer2023diffphase} & 30 & \U{4.04} & \U{0.97} & \U{-25.3} & \U{4.29} & \U{3.19} & \U{3.96} & \B{2.6\%} & \B{0.53} & $\infty$ & $\infty$ & 65.6M \\
        \bottomrule
    \end{tabular}
    \end{subtable}
    \hfill
    \begin{subtable}[t]{\textwidth}
    \centering
    \vspace{1em}
    \caption{Mel vocoding task. \emph{HiFi-GAN} refers to the 16\,kHz model trained on LibriTTS data \cite{zen2019libritts}, available in SpeechBrain \cite{ravanelli2024speechbrain}.
    We refer the reader to our prior work \cite{welker2025realtime} for a more in-depth evaluation on this task.}
    \label{tab:streamflow-melvocoder}
    \begin{tabular}{lrrrrrrrrrrrr}
        \toprule
        \B{Method} & \B{NFE} & \B{PESQ} & \B{ESTOI} & \B{SI-SDR} & \B{DiMOS} & \B{WVMOS} & \B{NISQA} & \B{WER$\downarrow$} & \B{LSD$\downarrow$} & $\latalg$ & $\lattot$ & \B{Params} \\
        \midrule
        $M^\dagger$ + Zero-phase & - & 1.28 & 0.63 & -38.9 & 1.43 & 1.07 & 1.21 & 35.1\% & 2.22 & - & - & - \\
        $M^\dagger$ + RTISI-DM \cite{peer2024flexible} & 50 & 2.97 & 0.88 & -29.7 & 2.51 & 1.92 & 2.58 & 5.8\% & 0.82 & 32 & 48 & 1 \\
        SFM Euler1 & 1 & 1.35 & \R{0.36} & \B{-5.6} & \R{1.18} & 1.23 & 1.35 & \R{86.7\%} & \R{7.82} & 32 & 48 & 27.9M \\
        SFM Euler5 & 5 & 4.10 & \B{0.96} & \U{-10.1} & \U{4.31} & \U{3.11} & 4.14 & 4.7\% & 1.01 & 32 & 48 & 27.9M \\
        SFM Midpoint2 & 4 & 3.92 & 0.94 & -10.2 & 4.28 & 2.99 & \B{4.17} & \U{4.5\%} & 0.89 & 32 & 48 & 27.9M \\
        SFM LRK5 \eqref{eq:custom-rk-loss} & 5 & \B{4.14} & \B{0.96} & -10.4 & \B{4.34} & \B{3.15} & \U{4.15} & \B{3.6\%} & \B{0.80} & 32 & 48 & 27.9M \\
        \midrule
        FM Euler5 & 5 & \B{4.34} & \B{0.97} & \B{-9.9} & \B{4.35} & \B{3.05} & \B{4.15} & \B{4.3\%} & \U{1.04} & $\infty$& $\infty$ & 38.7M \\
        HiFi-GAN \cite{kong2020hifigan} & 1 & \U{2.99} & \U{0.90} & \U{-29.9} & \U{4.21} & \U{3.02} & \U{3.91} & \U{5.4\%} & \B{0.77} & 236 & 252 & 13.9M \\
        \bottomrule
    \end{tabular}
    \end{subtable}
\end{table*}

\subsection{Computations, runtimes, and memory}\label{sec:results:timings-flops-latencies}
\begin{table}
    \centering
    \caption{FLOP and RTF measurements on an RTX 4080 Laptop GPU. $N$ is the number of \ac{dnn} calls per frame.
    FLOPs are per-frame, multiplied with the number of frames per second, streaming RTF is relative to each model's per-frame runtime budget (frame shift), and offline RTF is determined for 1-second inputs.
    Model variants as in \cref{tab:streamflow-se}.
    }
    \label{tab:flops-and-timings}
    \begin{tabular}{lrrr}
        \toprule
        \B{Method} & \B{GFLOPs/sec.} & \B{Streaming RTF} & \B{Offline RTF} \\
        \midrule
        Stream.FM & $N \cdot 282.0$ & $N \cdot 0.174$ & $N \cdot 0.021$ \\
        Diffusion Buffer \cite{lay2025diffusionjpre} & 3529.8 & 0.475 & 0.007 \\
        Causal SEMamba \cite{semamba} & 42.5 & 0.867 & 0.016 \\
        DeepFilterNet3 \cite{schroeter2023deepfilternet3} & 0.3 & 0.321 & 0.008 \\
        DEMUCS \cite{defossez2020realtime} & 10.2 & 0.221 & 0.004 \\
        CleanUMamba \cite{groot2025cleanumamba} & \emph{n/a} & 0.278 & 0.004 \\
        HiFi-Stream \cite{dmitrieva2025hifi} & 6.7 & 0.098 & 0.012 \\
        \bottomrule
    \end{tabular}
    \vspace{0em}
\end{table}
In \cref{tab:flops-and-timings} we list, for our model and various streaming-capable baselines: \B{(1)} the determined GFLOPs per second, \B{(2)} the streaming \ac{rtf}, defined as $\frac{t_{\mathrm{proc,fr}}}{\hopsynth/f_s}$ for each single frame where $f_s$ is the sampling frequency and $t_{\mathrm{proc,fr}}$ is the processing time per frame \cite{defossez2020realtime}, and \B{(3)} the offline RTF $\frac{t_{\mathrm{proc}}}{1\,\text{sec}}$, where $t_\mathrm{proc}$ is the time taken to process one 1-second utterance in a single model call.
We can see that offline \ac{rtf} consistently and severely underestimates the streaming \ac{rtf}, and that FLOPs have no simple relation to streaming \ac{rtf}, e.g., the Diffusion Buffer \cite{lay2025diffusionjpre} attains smaller streaming and offline RTFs than Stream.FM for 5 \ac{nfe} but spends substantially more FLOPs. For CleanUMamba \cite{groot2025cleanumamba}, no FLOPs measurement is available due to an incompatibility with the utilized FLOP counting method, see \cref{sec:eval:runtime-performance}, and we instead refer the reader to the GMACs values reported in the paper \cite{groot2025cleanumamba}.

On a modern CPU (AMD Ryzen 7 9800X3D), we measure the streaming RTF of Stream.FM as $N\times0.883$, which still permits real-time streaming with $N=1$ DNN evaluation and no initial predictor DNN, but precludes the use of any more than one model call, motivating the future search for more efficient DNN architectures.

Note that multiplying each model's streaming RTF by its frame shift yields the effective processing latency $\latproc$, and adding this to the algorithmic latency $\latalg$ provides a lower bound for each model's total latency $\lattot$, \emph{on this concrete hardware}. However, as long as the streaming RTF is below 1, simply using $\lattot = \winlensynth + \hopsynth$ is closer to typical frame-by-frame implementations, see \cref{sec:latency-definitions}.

Regarding memory usage in a real-time streaming setting, we analyze a streaming implementation of our \ac{se} model. We measure the total used GPU memory as $583 + (45 \cdot N)$ MiB. We further measure the memory used only by the tensors caching past activations, see \cref{sec:efficient-rt-diffusion-inference}, as 15 MiB per solver step (plus 15 MiB for the initial predictor). These values permit deployment on any modern consumer PC, and potentially also some embedded devices.

\subsection{Learned Runge-Kutta solvers}\label{sec:lrk-solvers-analysis}
\begin{figure}
    \centering
    \includegraphics[width=\linewidth]{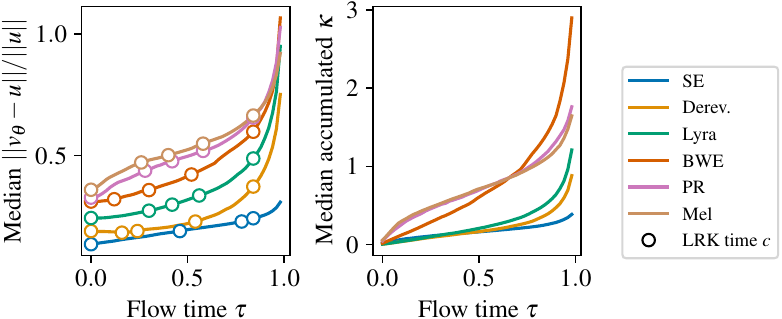}
    \caption{
    Model error and curvature of the learned models in each task. \B{Left:} Normalized model error compared to the constant ground-truth velocity $u$ at each $\tau$, and learned Runge-Kutta solver time-points $\mathbf c$ learned with \eqref{eq:custom-rk-loss} for each task. \B{Right:} geometric curvature $\kappa$ \cite{hubbard1998vector} accumulated up until each indicated $\tau$. A fully straight velocity field would have zero curvature $\kappa$ even under (constant) model error.
    \vspace{-1.5em}
    }
    \label{fig:problem-curvature-analysis}
\end{figure}
\begin{table}
    \centering
    \caption{Changes in metrics for bandwidth extension by using an \ac{lrk} solver trained with loss \eqref{eq:custom-rk-loss} instead of the Euler solver, compared at equal NFE. Improved values by \ac{lrk} in \G{green}, worsened values in \R{red}, biggest improvement / smallest degradation in \B{bold}.
    }
    \label{tab:streamflow-bwe-lrk-differences}
    \resizebox{\linewidth}{!}{
    \begin{tabular}{crrrrrrr}
        \toprule
        {\scriptsize \B{NFE}} & {\scriptsize \B{PESQ}} & {\scriptsize \B{SI-SDR}} & {\scriptsize \B{DiMOS}} & {\scriptsize \B{WVMOS}} & {\scriptsize \B{NISQA}} & {\scriptsize \B{WER$\downarrow$}} & {\scriptsize \B{LSD$\downarrow$}} \\
        \midrule
        5 & \R{-0.35} & \R{-1.20} & \G{0.12} & \G{0.06} & \G{0.23} & \G{-1.8\%} & \G{-0.26} \\
        4 & \R{-0.34} & \R{-0.81} & \G{0.23} & \G{0.11} & \G{0.30} & \G{-1.8\%} & \G{-0.32} \\
        3 & \R{-0.47} & \R{-1.20} & \B{\G{0.35}} & \B{\G{0.16}} & \B{\G{0.44}} & \B{\G{-2.3\%}} & \B{\G{-0.41}} \\
        2 & \B{\R{-0.20}} & \B{\R{-0.20}} & \G{0.25} & \G{0.11} & \G{0.30} & \G{-0.2\%} & \G{-0.30} \\
        \bottomrule
        \vspace{-2em}
    \end{tabular}
    }
\end{table}
In our previous evaluations, we have seen that the \ac{lrk} solvers can boost output quality both in objective metrics and in a listening test, but to which degree depends on the task and the loss used to train the solver. The \ac{lrk} solvers are especially effective for \ac{bwe}, \ac{stft} phase retrieval, and Mel vocoding, but do not yield such strong improvements for \ac{se}, dereverberation, and Lyra codec artifact removal.
As a possible explanation for this inconsistent improvement, we propose that it is related to the curvature
of the learned velocity field $v_\theta$ along the flow time $\tau$.
To support this, we show an empirical estimation of the model error and the geometric curvature $\kappa$ \cite{hubbard1998vector} in \cref{fig:problem-curvature-analysis}, which we determined from 100 trajectories calculated with $N=50$ Euler steps with each trained model for each task, on samples from each respective training dataset, see \crefnobox{sec:traj-curvature-determination} of the supplementary material.
We plot the \emph{accumulated} curvature $\kappa$, i.e., $\int_0^\tau \kappa(t) \mathrm{d}t$, to show both local changes in curvature and an informative overall curvature at $\tau=1$. We also plot the evaluation time-points $\mathbf c$ from the \ac{lrk} solver learned for each task using \eqref{eq:custom-rk-loss}.
We can clearly see that the \ac{se} task has much lower curvature than all other tasks, reflecting the linear additive corruption model and explaining why even a single Euler step yields very good results here.
The tasks then increase in total curvature in the following order: dereverberation, codec artifact removal, Mel vocoding and \ac{stft} phase retrieval, and \ac{bwe}. Dereverberation and codec artifact removal have lower curvature and may have less inherent ambiguity than for instance \ac{bwe},
and they profit more from using the signal-fidelity oriented PESQ-based loss \eqref{eq:custom-rk-loss-pesq} to train the \ac{lrk} solver, see \cref{tab:streamflow-derev,tab:streamflow-lyra}.

For the LRK4 results on \ac{se}, shown in \cref{tab:streamflow-se,tab:vb-dmd-results}, we used a single set of \ac{lrk} solver parameters trained on \ac{ewv2} with our main \ac{ourmethod} model for SE. From these tables, we can see that these parameters generalize reasonably well to other models (16~ms \ac{ourmethod} variants) and datasets (\ac{vbdmd} benchmark). Regarding cross-task evaluation, a complete evaluation, conducted by running inference for every task and model with every task-mismatched set of solver parameters, can be found in \crefnobox{sec:cross-lrk-evaluation} of the supplementary material. The results suggest that \B{(1)} the \ac{lrk} solvers for \ac{se}, dereverberation, codec artifact removal and \ac{bwe} transfer reasonably well to other tasks within this group, \B{(2)} the aforementioned solvers do not transfer well to \ac{pr} and Mel vocoding, and \B{(3)} most interestingly, the solvers for \ac{pr} and Mel vocoding conversely seem to be most universally applicable.

Another interesting question is how helpful \ac{lrk} solvers are in settings with even lower \ac{nfe}. To investigate this, we train one solver for each $N \in \{2, 3, 4, 5\}$ for bandwidth extension, and compare each against the Euler solver at the same respective \ac{nfe}. Note here that $N=1$ yields the plain Euler scheme with no learnable parameters. We show the results in \cref{tab:streamflow-bwe-lrk-differences}, where we find that the biggest improvement over Euler occurs at $N=3$, though also at the cost of the largest decrease in PESQ.

Finally, regarding the time needed to training a \ac{lrk} scheme, we find that this scales linearly  with $N$, taking $2.3N$ hours for 25,000 on an NVIDIA RTX A6000. This yields around 11.5 hours for a 5-stage \ac{lrk} solver, substantially less than the time taken to train the flow model DNN (62 hours) on the same GPU.

\subsection{Model weight compression for Mel vocoding}
\begin{figure}
\centering
    \includegraphics[width=\linewidth]{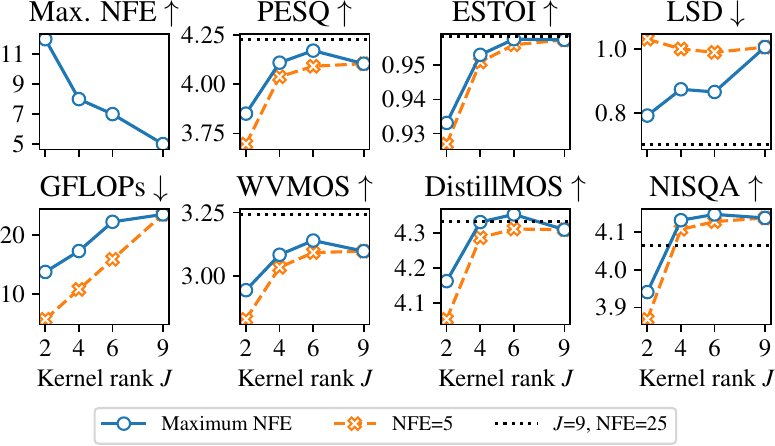}
\caption{%
Metrics of compressed Stream.FM models for Mel vocoding
using kernel ranks $J \in \{2,4,6,9\}$ where $J=9$ is uncompressed, using the Euler solver. We compare the maximum NFE for each $J$ under our runtime budget against constant $\text{NFE}=5$ and a high-NFE variant $K=9,\ \text{NFE}=25$. Reported GFLOPs are per-frame.
}
\label{fig:model-compression}
\end{figure}

In \cref{fig:model-compression}, we analyze the model weight compression ideas introduced in \cref{sec:model-compression}, using the Mel vocoding task as an example with the Euler solver at different \ac{nfe}. We can see that more compressed models (smaller ranks $T<K$) reduce all metrics when keeping NFE\,=\,5 constant, and linearly decrease GFLOPs per frame, both as expected. However, when increasing the NFE to the respective maximum possible number for each compressed model on our hardware, we find that $T=6$ at NFE\,=\,7 is better across all metrics than the uncompressed $T=K=9$ at NFE\,=\,5, while also slightly decreasing the GFLOPs per frame. This suggests that slight model compression may be preferable over no compression if it allows to increase the NFE, confirming the usefulness of our proposal.

\subsection{Model architecture ablations}\label{sec:ablation-study}
\begin{table}
    \centering
    \caption{
    DNN architecture ablation (EARS-Reverb v2). \B{C} indicates frame-causality.
    GFLOPs per second. \G{Better}/\R{worse} than SFM.
    Causal attention is unoptimized, hence true $\Delta$GFLOPs is lower.
    }
    \label{tab:dnn-ablation-results}
    \resizebox{\linewidth}{!}{
    \begin{tabular}{lrrrrl}
        \toprule
        \B{Change} & \B{$\Delta$GFLOPs} & \B{PESQ} & \B{DiMOS} & \B{WER}$\downarrow$ & \B{C} \\
        \midrule
        FM & -68.9 & 2.31 & 3.77 & 11.4\% & \xmark \\
        Stream.FM & 0.0 & 2.01 & 3.76 & 16.8\% & \cmark \\
        \arrayrulecolor{black!20}\midrule\arrayrulecolor{black}
        +Bottleneck attn. & \R{+5.2$N$} & \G{2.33} & \G{3.90} & \G{13.0\%} & \xmark \\
        +Causal win. bottl. attn. & \R{$\leq$ +5.2$N$} & \G{2.04} & \G{3.79} & \G{16.6\%} & \cmark \\
        +Concat. fusion & \R{+62.9$N$} & \R{2.00} & 3.76 & \R{17.5\%} & \cmark \\
        +Cum. GroupNorm & +0.0 & \G{2.06}  & 3.76 & \R{16.9\%} & \cmark \\
        2$\rightarrow$1 residual blocks & \G{-85.6$N$} & \R{1.81} & \R{3.58} & \R{20.7\%} & \cmark \\
        \bottomrule
        \vspace{-2.5em}
    \end{tabular}
    }
\end{table}
As guidance for future work, we conduct ablation studies for some of our architectural choices in \cref{sec:dnn-modifications}, modifying only one component of the \ac{dnn} at a time.
We choose dereverberation as the task, motivated by the assumption that it especially profits from effective modeling of intricate long-range and phase relationships. We investigate the addition of non-causal bottleneck attention, the addition of causal windowed bottleneck attention with 64 past frames (1.024 seconds), using channel-concatenation instead of addition for skip connections, using a cumulative GroupNorm as in \cite{richter2024causal} instead of our SGBatchNorm, and using one instead of two residual blocks per layer. We use the Euler solver with 5 steps for a consistent comparison, and also list the non-causal FM model for comparison.

We show the resulting PESQ, DistillMOS, and WER values in \cref{tab:dnn-ablation-results}, along with the change in GFLOPs per second from each modification. We see that adding non-causal bottleneck attention strongly boosts the output quality, and is even slightly better in PESQ and DistillMOS than the non-causal FM model, which also uses bottleneck attention but on time-downsampled features.
However, using causal windowed bottleneck attention instead to get a streaming-capable variant provides only a very minor quality increase. This supports our earlier claim that future information is very helpful in the dereverberation task, and justifies our choice of removing attention for simplicity and efficiency in our streaming-oriented setting.
Using channel concatenation for skip connections does not yield any improvement while requiring $62.9N$ more GFLOPs per second, justifying our simpler choice of addition.
Using a cumulative GroupNorm as in \cite{richter2024causal} instead of our SGBatchNorm slightly increases PESQ, but makes no clear overall positive or negative difference. We argue that in dynamic real-world acoustic environments (e.g., moving between rooms), a time-invariant norm such as ours is likely still preferable.
Using one residual block instead of two clearly decreases all three metrics, supporting our use of two residual blocks.

\section{Conclusion}
\label{sec:conclusion}
In this work, we have presented Stream.FM, a streaming generative method for general speech restoration tasks which can run in real-time on a consumer GPU. We detailed our architecture, inference scheme, and other optimizations needed to achieve this real-time capability, showed state-of-the-art performance for generative streaming methods through metrics and listening experiments, and proposed novel ideas to optimize quality or compute in a low-\ac{nfe} setting. Our contributions and public codebase aim at closing the gap between generative and predictive speech restoration models in real-time settings, and at supporting the research community towards further developments in this field.

\section{Acknowledgements}

We acknowledge funding by the German Federal Ministry of Research, Technology and Space (BMFTR) under grant agreement No.~16IS24072B (COMFORT); by the Deutsche Forschungsgemeinschaft (DFG, German Research Foundation) -- 498394658; and by the Federal Ministry for Economic Affairs and Climate Action (Bundesministerium für Wirtschaft und Klimaschutz), Zentrales Innovationsprogramm Mittelstand (ZIM), Germany, within the project FKZ KK5528802VW4.

\bibliographystyle{IEEEtran}
\bibliography{refs}

\begin{IEEEbiography}[{\includegraphics[width=1in,height=1.25in,clip,keepaspectratio]{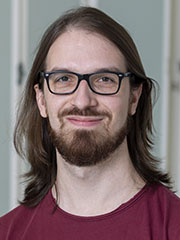}}]{Simon Welker}
received a B.Sc. in Computing in Science (2019) and M.Sc. in Bioinformatics (2021) from University of Hamburg, Germany. He is currently a PhD student in the labs of Prof. Timo Gerkmann (Signal Processing, University of Hamburg) and Prof. Henry N. Chapman (Center for Free-Electron Laser Science, DESY, Hamburg), researching machine learning techniques for solving inverse problems that arise in speech processing and X-ray imaging. He received the VDE ITG award 2024.
\end{IEEEbiography}%
\begin{IEEEbiography}[{\includegraphics[width=1in,height=1.25in,clip,keepaspectratio]{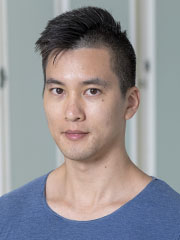}}]{Bunlong Lay}
obtained a B.Sc. and M.Sc. in Mathematics in 2015 and 2017 from the University of Bonn, Germany. He subsequently joined the research institute 
Fraunhofer FKIE in Wachtberg Germany from 2018 until 2021, where he focused on research in the field of radar signal processing. In 2021 he started his Ph.D. at the University of Hamburg. Currently researching Diffusion-based models for Speech Enhancement for real-time applications. He received the VDE ITG award 2024.
\end{IEEEbiography}%
\begin{IEEEbiography}[{\includegraphics[width=1in,height=1.25in,clip,keepaspectratio]{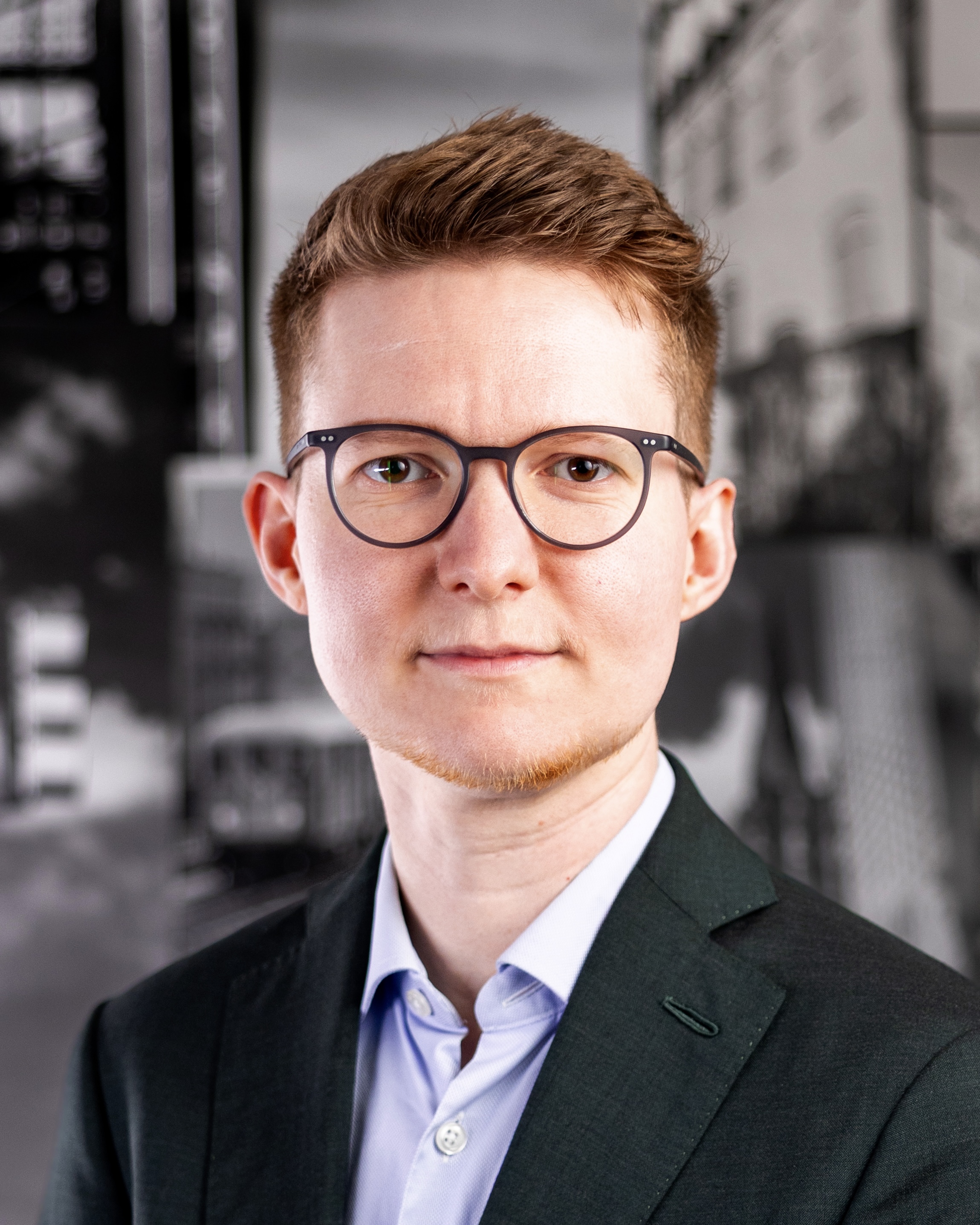}}]{Maris Hillemann}
obtained a B.Sc. in Computer Science in 2024 from the University of Hamburg, Germany. He is currently a master's student in Computer Science at the University of Hamburg and a student assistant in the Signal Processing group of Prof. Timo Gerkmann.
\end{IEEEbiography}%
\begin{IEEEbiography}[{\includegraphics[width=1in,height=1.25in,clip,keepaspectratio]{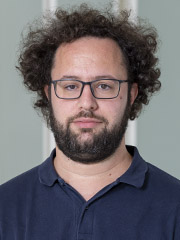}}]{Tal Peer} received the B.Sc. degree in General Engineering Science (2016) and the M.Sc. degree in Electrical Engineering (2019) from the Hamburg University of Technology. He is currently pursuing a PhD with the Signal Processing group at the University of Hamburg. His research interests include phase-aware speech enhancement and phase retrieval for speech and audio applications.
\end{IEEEbiography}%
\begin{IEEEbiography}[{\includegraphics[width=1in,height=1.25in,clip,keepaspectratio]{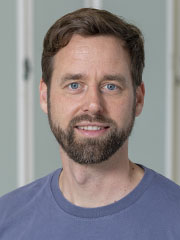}}]{Timo Gerkmann}
(S’08–M’10–SM’15) is a professor for Signal Processing at the Universität Hamburg, Germany. He has previously held positions at Technicolor Research \& Innovation in Germany, the University of Oldenburg in Germany, KTH Royal Institute of Technology in Sweden, Ruhr-Universität Bochum in Germany, and Siemens Corporate Research in Princeton, NJ, USA. His main research interests are on statistical signal processing and machine learning for speech and audio applied to communication devices, hearing instruments, audio-visual media, and human-machine interfaces. Timo Gerkmann served as member of the IEEE Signal Processing Society Technical Committee on Audio and Acoustic Signal Processing and is currently a Senior Area Editor of the IEEE/ACM Transactions on Audio, Speech and Language Processing. He received the VDE ITG award 2022.
\end{IEEEbiography}

\vfill  %

\clearpage
\section*{Supplementary Material}
\renewcommand{\thefigure}{S\arabic{figure}}
\renewcommand{\thetable}{S\Roman{table}}
\renewcommand{\thesection}{S.\Roman{section}}

\section{Model implementation and optimization}\label{sec:model-impl-and-optimizations}
In this supplementary material, we list additional technical details for our Stream.FM models, which are not necessary for the method itself but are useful to implement real-time capable streaming inference.

\subsection{Functional frame-wise inference API}
To implement our frame-wise inference scheme, we implement our model using PyTorch and PyTorch Lightning.
We design an \ac{api} similar to the one proposed for CINs \cite{hedegaard2022cins}. The official CINs implementation stores tracked states on each layer, making every layer stateful. This is problematic for our purposes, since we want to use only a single \ac{dnn} instance $\lvel$ with shared weights, but multiple state collections (one for each diffusion solver step).

Our \ac{api} hence aims to make the layers themselves stateless, so they do not track their own state. It is based on two functions, \texttt{init\_state()} and \texttt{forward\_step(x, state)}, which each stateful streaming layer and the overall model must implement. \texttt{init\_state()} creates and returns a data structure containing all tracked state variables (e.g., buffers) of each module as well as their respective nested modules. \texttt{forward\_step(frame, state)} receives a single new \texttt{frame} as well as the previous \texttt{state}, and returns an output frame along with an updated \texttt{state} data structure. Higher-level modules then receive, pass along, and update their own state as well as the state for any nested modules.

We note that, unlike pure functional programming \acp{api}, our \ac{api} does not necessarily construct a new \texttt{state} object for each call, but may instead modify an existing \texttt{state} in-place. We made this choice for better compatibility with \texttt{torch.compile} and CUDA graphs, but leave further optimization to future work.

\subsection{Minimizing overhead}
In initial experiments with our models for real-time streaming inference, we found that the processing time per frame quickly exceeded our 16\,ms budget even for low \ac{nfe} values such as 3. After careful inspection of profiler traces, we found that the processing was dominated by CPU-GPU overhead. To reduce this overhead as much as possible, we first tried using model compilation via \texttt{torch.compile} to merge CUDA kernels, which reduced the overhead to some extent but not enough to allow real-time inference. We then modified our code to use CUDA Graphs as implemented in PyTorch. By capturing the entire solver computation including the sequence of $N$ DNN calls in a single CUDA Graph and then replaying this CUDA Graph for every frame, we are able to realize up to 5 DNN calls per frame within the 16\,ms time budget.

\section{Further details on learned Runge-Kutta solvers}
We provide further details on our proposed \ac{lrk} solvers in the following.

\subsection{Numerical properties of our learned RK solvers}
Due to the constraints we impose on the learned \ac{rk} coefficients, namely $\sum_j a_{ij} = c_i, \sum_i b_i = 1$ as described in the main paper, all our \ac{lrk} methods are first-order consistent and have order of convergence of at least 1, by construction. Furthermore, as (consistent) Runge-Kutta schemes, they are (consistent) one-step methods and thus automatically have the property of zero-stability \cite{leveque2007finite}.

Through inspection of the learned coefficients, we find that higher convergence orders than 1 are not attained. This is expected, as such conditions require very specific algebraic relationships between the coefficients \cite{hairer1993solving} that are extremely unlikely to be found by any learning-based method of determining the coefficients. This is, however, also acceptable since the solvers are intended to run at our fixed \ac{nfe} budget and therefore at a fixed step $\Delta t$ (particularly in our case $\Delta t = 1$), and are not intended to be general-purpose solvers.

We note here that, unlike the more general usage of \ac{rk} \ac{ode} solvers which are called multiple times in sequence with some choice of step size $\Delta\tau$, our solvers are trained with $\Delta\tau = 1$ for the fixed time horizon $\tau \in [0, 1]$, and are intended to be called only once for each frame in order to guarantee a fixed \ac{nfe} within our streaming constraints. Furthermore, the \ac{ode} induced by the learned flow matching velocity model $\lvel$ is extremely high-dimensional and generally non-linear. This means that solver stability properties, which are based on the analysis of a low-dimensional linear test \ac{ode} and usually relate to long-term stability of repeated solver calls over unbounded time intervals \cite{butcher2016numericalode}, are less relevant for our single-call solvers than in a classical numerical setting.

\begin{figure}
    \centering
    \includegraphics[width=\linewidth]{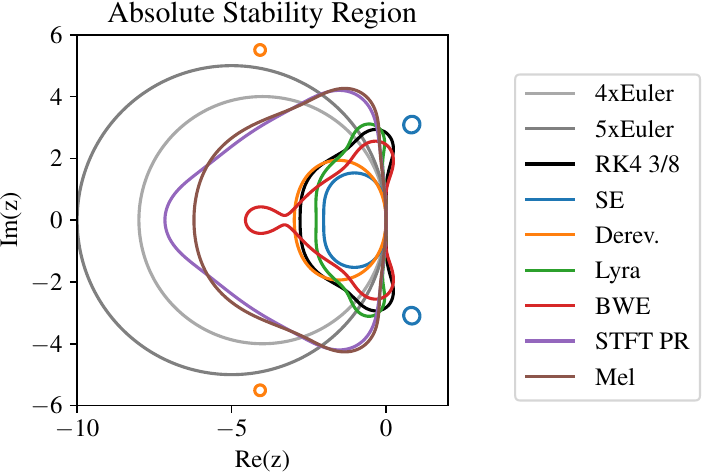}
    \caption{Regions of absolute stability for each \acf{lrk} solver listed in \cref{sec:learned-rk-coefs}, determined from the coefficients $\mathbf A, \mathbf b$ as in \cite{butcher2016numericalode}, see \cref{eq:rk-stability-function}. For comparison, we also plot four and five Euler steps -- each treated as one Runge-Kutta scheme -- and the original fourth-order Runge-Kutta 3/8 scheme.}
    \label{fig:lrk-absolute-stability-regions}
\end{figure}

Nonetheless, in line with classic literature on numerical \ac{ode} solvers \cite{butcher2016numericalode}, we can analyze and plot the \emph{region of absolute stability} of each of our schemes, which can help characterize the solvers and may hold insights to be explored in future works. To this end, we introduce the stability function for Runge-Kutta schemes as in \cite{butcher2016numericalode}:
\begin{equation}\label{eq:rk-stability-function}
    R(z) := 1 + z \mathbf{b}^\top
(I-z \mathbf{A})^{-1} \mathbf{1}
\end{equation}
where $\mathbf A, \mathbf b$ are taken from the coefficients of the \ac{lrk} scheme and $\mathbf 1$ indicates the vector $[1, \ldots, 1]$. The region of absolute stability is then the region of the complex plane where the condition $|R(z)| < 1$ holds. We plot these regions in \cref{fig:lrk-absolute-stability-regions} for all \ac{lrk} schemes learned with the SpeechBERTScore-based loss. The particularly large regions of absolute stability of the \ac{lrk} solvers for the STFT phase retrieval and Mel vocoding tasks is interesting, and may relate to their high reliability across tasks which we find in \cref{sec:cross-lrk-evaluation}. We reserve deeper investigations of possible relations for future work.

\subsection{Learned coefficients for learned RK solvers}\label{sec:learned-rk-coefs}
In this section, we list the parameters $\{\mathbf A, \mathbf b, \mathbf c\}$ of the learned Runge-Kutta \ac{ode} solver schemes, see Section III.E in the main paper, for each of the six tasks investigated in the main paper and for every solver learned with the SpeechBERTScore-based loss, Eq. (19) in the main manuscript. Note that values are rounded to three decimal points for clarity, and that the learned values at full precision always ensure the conditions we impose as described in the main paper (up to numerical floating-point error), e.g., $\sum_i b_i = 1$.

\subsection{Speech enhancement}
\begin{equation}\label{eq:lrk-params-se}
    \begin{aligned}
        \mathbf A &= \begin{bmatrix}
        0 & 0 & 0 & 0 \\
        0.458 & 0 & 0 & 0 \\
        -0.847 & 1.623 & 0 & 0 \\
        2.029 & -1.707 & 0.528 & 0
        \end{bmatrix} \\
        \mathbf b &= \begin{bmatrix}
        0.339, & 0.444, & 0.102, & 0.114
        \end{bmatrix} \\
        \mathbf c &= \begin{bmatrix}
        0, & 0.458, & 0.776, & 0.850
        \end{bmatrix} \\
    \end{aligned}
\end{equation}

\subsection{Dereverberation}
\begin{equation}\label{eq:lrk-params-derev}
    \begin{aligned}
        \mathbf A &= \begin{bmatrix}
        0 & 0 & 0 & 0 & 0 \\
        0.152 & 0 & 0 & 0 & 0 \\
        -0.065 & 0.312 & 0 & 0 & 0 \\
        0.088 & 0.296 & 0.152 & 0 & 0 \\
        0.565 & 0.856 & 1.425 & -1.997 & 0
        \end{bmatrix} \\
        \mathbf b &= \begin{bmatrix}
        0.079, & 0.223, & 0.423, & 0.184, & 0.091
        \end{bmatrix} \\
        \mathbf c &= \begin{bmatrix}
        0, & 0.152, & 0.247, & 0.536, & 0.850
        \end{bmatrix} \\
    \end{aligned}
\end{equation}

\subsection{Codec post-filtering}
\begin{equation}\label{eq:lrk-params-lyra}
    \begin{aligned}
        \mathbf A &= \begin{bmatrix}
        0 & 0 & 0 & 0 & 0 \\
        0.298 & 0 & 0 & 0 & 0 \\
        0.049 & 0.375 & 0 & 0 & 0 \\
        -0.245 & 1.030 & -0.219 & 0 & 0 \\
        0.672 & -0.168 & -0.276 & 0.622 & 0
        \end{bmatrix} \\
        \mathbf b &= \begin{bmatrix}
        0.089, & 0.211, & 0.307, & 0.100, & 0.292
        \end{bmatrix} \\
        \mathbf c &= \begin{bmatrix}
        0, & 0.298, & 0.424, & 0.566, & 0.850
        \end{bmatrix} \\
    \end{aligned}
\end{equation}

\subsection{Bandwidth extension}
\begin{equation}\label{eq:lrk-params-bwe}
    \begin{aligned}
        \mathbf A &= \begin{bmatrix}
        0 & 0 & 0 & 0 & 0 \\
        0.112 & 0 & 0 & 0 & 0 \\
        -0.244 & 0.535 & 0 & 0 & 0 \\
        -1.093 & 1.840 & -0.217 & 0 & 0 \\
        -1.587 & 1.783 & 0.236 & 0.419 & 0
        \end{bmatrix} \\
        \mathbf b &= \begin{bmatrix}
        0.085, & 0.211, & 0.262, & 0.097, & 0.344
        \end{bmatrix} \\
        \mathbf c &= \begin{bmatrix}
        0, & 0.112, & 0.291, & 0.529, & 0.850
        \end{bmatrix} \\
    \end{aligned}
\end{equation}

\subsection{STFT phase retrieval}
\begin{equation}\label{eq:lrk-params-stft-pr}
    \begin{aligned}
        \mathbf A &= \begin{bmatrix}
        0 & 0 & 0 & 0 & 0 \\
        0.271 & 0 & 0 & 0 & 0 \\
        0.216 & 0.198 & 0 & 0 & 0 \\
        -0.029 & 0.147 & 0.454 & 0 & 0 \\
        0.072 & 0.208 & 0.326 & 0.244 & 0
        \end{bmatrix} \\
        \mathbf b &= \begin{bmatrix}
        0.128, & 0.209, & 0.307, & 0.130, & 0.227
        \end{bmatrix} \\
        \mathbf c &= \begin{bmatrix}
        0, & 0.271, & 0.413, & 0.572, & 0.850
        \end{bmatrix} \\
    \end{aligned}
\end{equation}

\subsection{Mel vocoding}
\begin{equation}\label{eq:lrk-params-mel-voc}
    \begin{aligned}
        \mathbf A &= \begin{bmatrix}
        0 & 0 & 0 & 0 & 0 \\
        0.251 & 0 & 0 & 0 & 0 \\
        0.104 & 0.286 & 0 & 0 & 0 \\
        -0.005 & 0.200 & 0.379 & 0 & 0 \\
        0.091 & 0.181 & 0.344 & 0.234 & 0
        \end{bmatrix} \\
        \mathbf b &= \begin{bmatrix}
        0.134, & 0.208, & 0.307, & 0.122, & 0.229
        \end{bmatrix} \\
        \mathbf c &= \begin{bmatrix}
        0, & 0.251, & 0.390, & 0.574, & 0.850
        \end{bmatrix} \\
    \end{aligned}
\end{equation}

\subsection{Mismatched cross-task evaluation of learned RK solvers}\label{sec:cross-lrk-evaluation}
\begin{figure*}
    \centering
    \includegraphics[width=.9\linewidth]{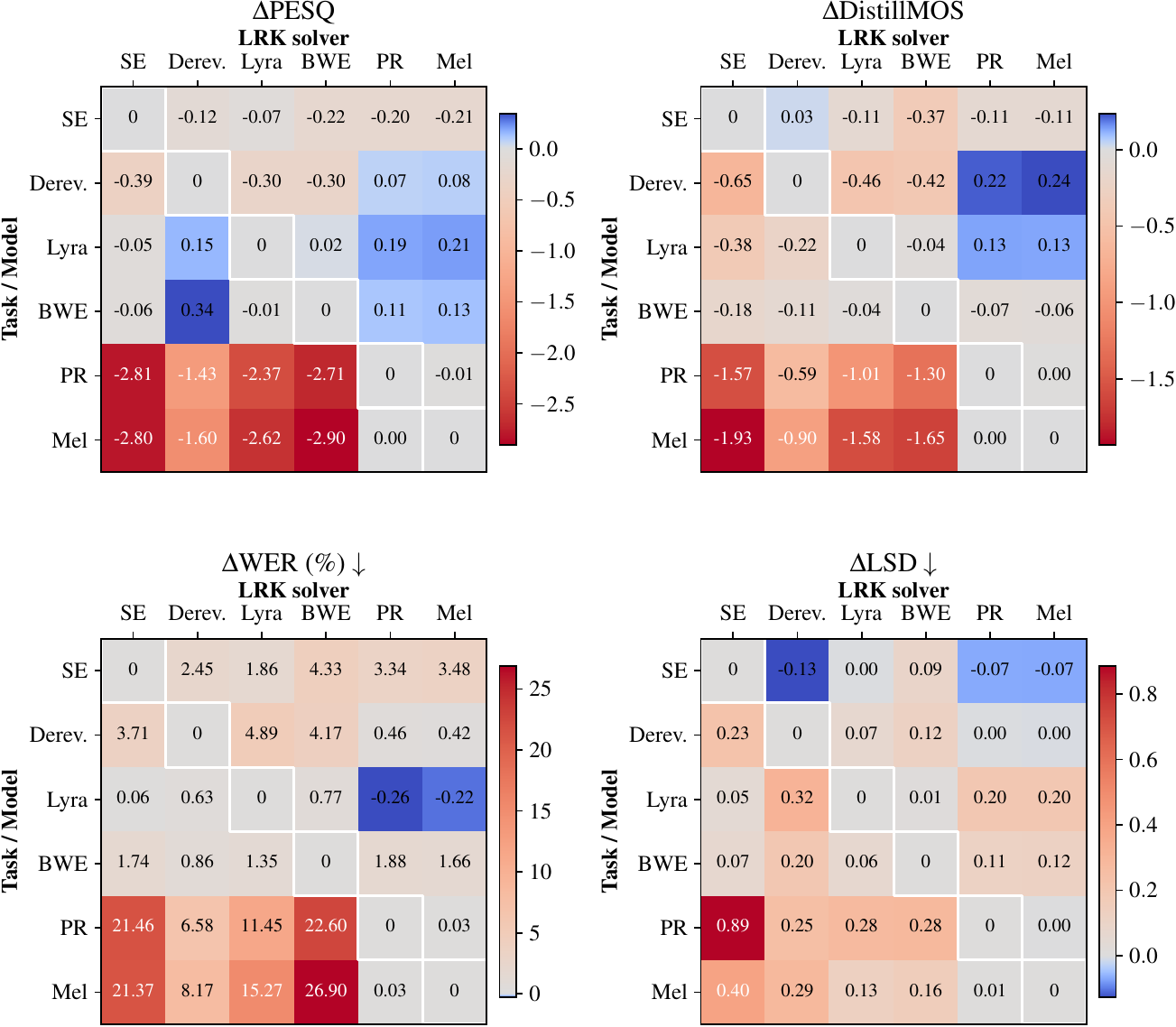}
    \caption{Cross-task evaluation of the different learned Runge-Kutta solvers, shown as differences $\Delta$ compared to the matched task-solver pairing.
    The \B{rows} indicate the task and the corresponding pretrained model, and the \B{columns} indicate that we used the Runge-Kutta solver with the coefficients originally fitted for the listed task, see \cref{sec:learned-rk-coefs}.
    Red indicates a worsening of the respective metric, whereas blue indicates an improvement.
    For WER and LSD, lower is better, hence the color-maps are flipped.
    Note that the solver for SE has only 4 stages, while all other solvers use 5 stages.
    }
    \label{fig:cross-lrk-evaluation}
\end{figure*}

In \cref{fig:cross-lrk-evaluation}, we show the results of an experiment where \enquote{mismatched} learned Runge-Kutta (LRK) solvers are used together with a model trained for a different task. Concretely, for each task we take the main trained Stream.FM model for this task from the main paper along with the corresponding test dataset, and then run inference with this model using each of the LRK solver parameters listed in \cref{sec:learned-rk-coefs} that were \B{not} trained for this task. This forms the Cartesian product
\begin{equation*}
    \left\{(t_i, s_j)\ |\ t_i \in T \wedge s_j=(\mathbf A^j, \mathbf b^j, \mathbf c^j) \in S \wedge i \neq j\right\}
\end{equation*}
where $T$ is the set of all original (task, model, dataset) pairs and $S$ is the set of corresponding learned RK solvers listed in \cref{sec:learned-rk-coefs}. For each such mismatched combination, we then calculate the empirical mean difference (improvement / degradation) compared to the matched pairing $(t_i, s_i)$, and plot the results in \cref{fig:cross-lrk-evaluation}.

Repeating the statement made on this evaluation in the main paper, \enquote{these results suggest that \B{(1)} the LRK solvers for \ac{se}, dereverberation, codec artifact removal and \ac{bwe} transfer reasonably well to other tasks within this group, \B{(2)} the aforementioned solvers do not transfer well to the \ac{pr} and Mel vocoding tasks, and \B{(3)} most curiously, the solvers learned for \ac{pr} and Mel vocoding seem to be most universally applicable, even sometimes improving some metrics compared to solvers matched with their target tasks}. Additionally, we note here the observations that: \B{(4)} the SE solver seems least widely applicable, and conversely the SE task seems to generally suffer most from using solvers trained for other tasks. This may have to do with the uniquely low velocity field curvature of the SE task compared to all other tasks, see the main paper; \B{(5)} the Lyra model seems to work better with the PR or Mel solvers than with the solver trained for this very task, though for a small increase in LSD. This further supports our observed difficulty of training an LRK solver for this specific task.

\section{Trajectory curvature determination}\label{sec:traj-curvature-determination}
In the main manuscript, we used the geometric curvature $\kappa$ \cite{hubbard1998vector} as a quantity to investigate a possible relationship between \B{(a)} the curvature of the solution trajectories associated with each specific speech restoration task and \B{(b)} the quality gains observed from the \acf{lrk} solvers over the plain Euler solver. We provide the details on how we determined the quantity $\kappa$ from sampled solution trajectories in the following, as they would exceed the scope of the main paper. The goal is to approximate the continuous notion of curvature $\kappa$ at time $\tau$ in our discretized setting,
\begin{equation}
    \kappa(\tau) = \left\lVert \frac{\mathrm{d}\mathbf{T}(\tau)}{\mathrm{d}s} \right\rVert
\end{equation}
where $\mathbf{T}(\tau) = \frac{v_t}{\lVert v_t \rVert}$ denotes the unit tangent vector along the trajectory curve $x_\tau$ at time $\tau$, and $s$ is the arc length along the curve for which it holds that $\frac{\mathrm{d}s}{\mathrm{d}\tau} = \lVert v_\tau \rVert$.
Using the chain rule for a parameterization in terms of $\tau$, we have
\begin{equation}
    \frac{\mathrm{d}\mathbf{T}}{\mathrm{d}s} = \frac{\mathrm{d}\mathbf{T}/\mathrm{d}\tau}{\mathrm{d}s/\mathrm{d}\tau} = \frac{\mathbf T'(\tau)}{\lVert v_\tau \rVert}
\end{equation}
where the $'$ indicates differentiation with respect to $\tau$. Therefore
\begin{equation}\label{eq:kappa-tprime}
    \kappa(\tau) = \frac{\lVert \mathbf T'(\tau) \rVert}{\lVert v_\tau \rVert},
\end{equation}
and we can determine $\mathbf T'$ using the quotient rule as
\begin{equation}\label{eq:curvature-calculation}
\begin{aligned}
    \mathbf T'(\tau) &= \frac{\mathrm{d}}{\mathrm{d}\tau} \frac{v_\tau}{\lVert v_\tau \rVert}
    = \frac{v_\tau' \lVert v_\tau \rVert - v_\tau \left( \frac{\mathrm{d}}{\mathrm{d}\tau} \lVert v_\tau \rVert \right)}{ \lVert v_\tau \rVert^2}  \\
    &= \frac{v_\tau' \lVert v_\tau \rVert - v_\tau \frac{\langle v_\tau', v_\tau \rangle}{\lVert v_t \rVert} }{ \lVert v_\tau \rVert^2} = \frac{a_\tau - \frac{\langle a_\tau, v_\tau \rangle}{\lVert v_t \rVert^2} v_\tau }{ \lVert v_\tau \rVert} \\
    &= \frac{a_{\tau,\perp}}{\lVert v_\tau \rVert}
\end{aligned}
\end{equation}
where $v_\tau$ is the velocity along the curve, $a_\tau$ is the acceleration, and
\begin{equation}
a_{\tau,\perp} := a_\tau - \frac{\langle a_\tau, v_\tau \rangle}{\lVert v_t \rVert^2} v_\tau
\end{equation}
is the component of $a_\tau$ perpendicular to $v_\tau$ as given by a Gram-Schmidt orthogonalization. This is intuitively sensible: The part of the acceleration that points in the same direction as the already present velocity will not induce any curvature, and should therefore be subtracted. With \eqref{eq:kappa-tprime}, \eqref{eq:curvature-calculation} now provides an expression for $\kappa$ purely in terms of accelerations $a_\tau$ and velocities $v_\tau$ along the trajectory, which is viable to calculate in our setting.

To calculate an approximate $\kappa$ in our discretized setting, we use the learned flow models $\lvel$ for each task and the Euler solver at high NFE ($N=50$) to generate 100 solution trajectories with low approximation error, using 100 randomly sampled utterance pairs $(X,Y)$ from each task's corresponding training dataset. Our choice $N=50$ means we use a Euler solver step size of $\Delta\tau = 1/N = 1/50 = 0.02$. We then have a sequence of points $X_\tau$ and associated velocities $V_\tau = \lvel(\cdot, \tau)$ along each generated trajectory. From this, we estimate the local acceleration as a first-order forward finite difference $A_t = \frac{V_{\tau+1} - V_\tau}{\Delta\tau}$.

After calculating $\kappa$ for every trajectory and every task, we aggregate it into a \emph{accumulated} curvature along each trajectory, $\int_0^\tau \kappa(t) dt$. This quantity provides both a local picture -- the increase in curvature visible at each $\tau$ -- and a global picture -- the overall curvature along the entire path so far, in each task. A single scalar that quantifies the total curvature can then be found at the accumulated curvature at $\tau=1$. Since the curvature and accumulated curvature differ between every trajectory, in order to provide a straightforward curvature comparison between tasks, we calculate the empirical population median of the accumulated curvatures at every $\tau$ over all 100 samples in each task, and plot the median accumulated curvatures over $\tau$.

\section{SNR-split metric evaluation on EARS-WHAM v2}
\begin{figure*}
    \centering
    \includegraphics[width=\linewidth]{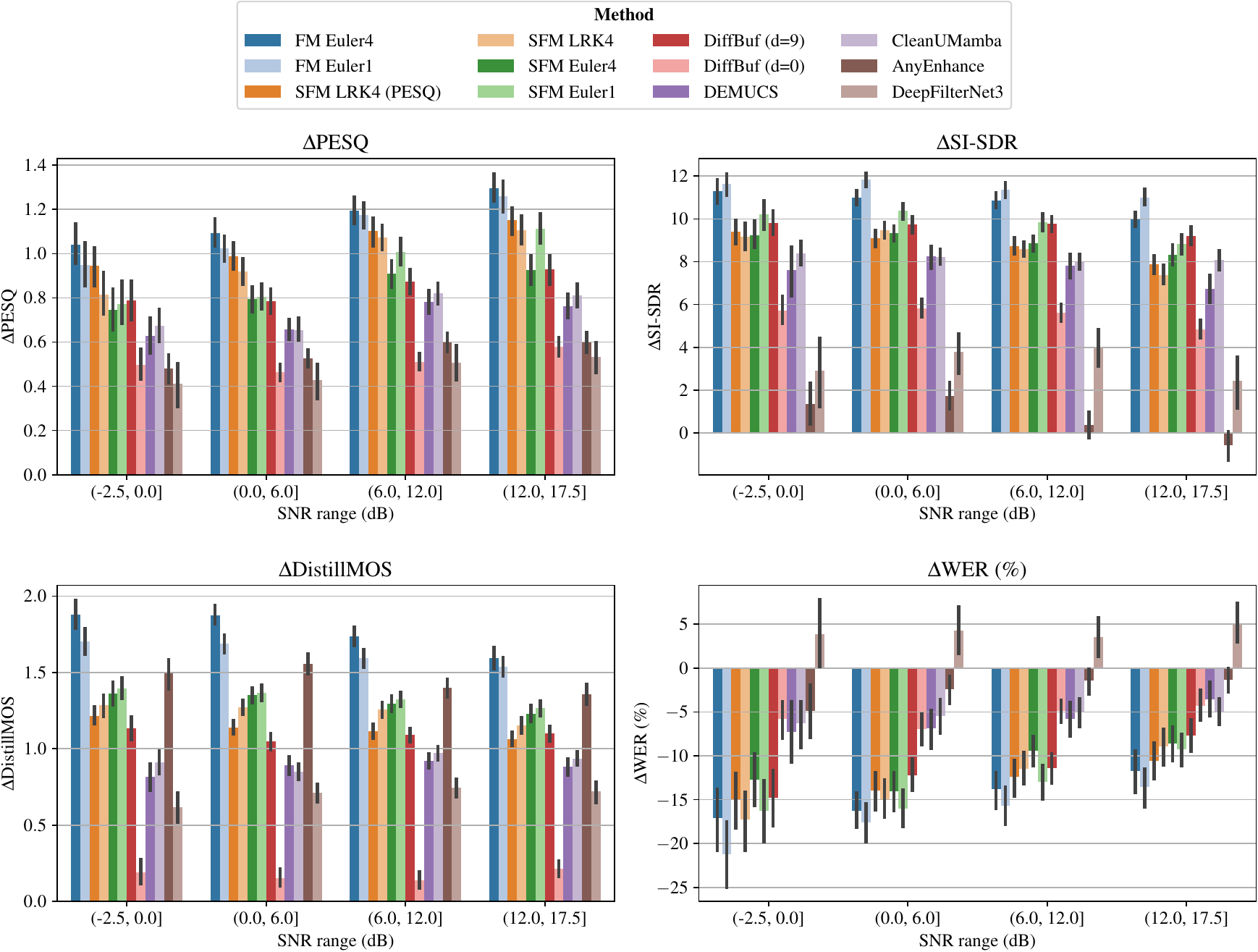}
    \caption{Speech enhancement task on the EARS-WHAM v2 test set: Metric improvements over the noisy inputs per input SNR bin, plotting the change $\Delta$ in PESQ, SI-SDR, DistillMOS, and WER for each listed method. We split the SNR (in dB) into four bins: [-2.5, 0], [0, 6], [6, 12], and [12, 17.5]. Small gray bars indicate the 95\% confidence interval.}
    \label{fig:snr-split-evaluation}
\end{figure*}

See \cref{fig:snr-split-evaluation} for a metric evaluation on the EARS-WHAM v2 test set which is split into four input SNR bins. For each SNR bin, we plot the improvement/degradation relative to the noisy inputs in four metrics (PESQ , SI-SDR, DistillMOS, \ac{wer}) for the most relevant \ac{se} models listed in the main paper. We find a very consistent behavior of all methods across the different SNR regions, with their relative ranks within each bin mostly preserved for each metric.

\end{document}